\documentclass[twocolumn,amsmath,amssymb,nofootinbib,superscriptaddress]{revtex4-2}

\usepackage{graphics,amssymb,amsmath,epsfig,color,textgreek}
\usepackage{graphicx,soul}
\usepackage[dvipsnames]{xcolor}
\usepackage{dcolumn}
\usepackage{bm}
\usepackage[colorlinks=true,citecolor=cyan]{hyperref}
\hypersetup{colorlinks=true,citecolor=cyan,linkcolor=red,urlcolor=magenta}
\usepackage{braket}
\usepackage[normalem]{ulem}
\usepackage{cancel}
\usepackage{xfrac}
\usepackage{amsfonts}
\usepackage{amsmath}
\usepackage{amssymb}
\usepackage{physics}
\usepackage{svg}
\usepackage{ORCIDinREVTeX}
\usepackage{lipsum}
\usepackage{thmtools}
\usepackage{thm-restate}

\usepackage[capitalise]{cleveref}
\crefname{section}{Sec.}{Secs.}
\crefname{table}{Tab.}{Tabs.}
\crefname{figure}{Fig.}{Figs.}
\crefname{definition}{Def.}{Defs.}
\crefname{lemma}{Lem.}{Lems.}
\crefname{theorem}{Thm.}{Thms.}
\crefname{corollary}{Cor.}{Cors.}
\crefname{adefinition}{Def.}{Defs.}
\crefname{alemma}{Lem.}{Lems.}
\crefname{atheorem}{Thm.}{Thms.}
\crefname{acorollary}{Cor.}{Cors.}

\renewcommand{\H}{\mathbb{H}}

\DeclareMathOperator{\SVT}{SVT}

\DeclareMathOperator{\proj}{proj}
\DeclareMathOperator{\vectorize}{vec}

\DeclareMathOperator{\divword}{div}
\DeclareMathOperator{\PSVT}{PSVT}
\DeclareMathOperator{\NMSE}{NMSE}

\newcommand{\R}{\mathbb{R}}
\newcommand{\C}{\mathbb{C}}

\begin{document}

\title{Approximate Message Passing for Quantum State Tomography} 
\author{Noah Siekierski}
\orcid{0000-0003-3300-3008}
\affiliation{Department of Electrical and Computer Engineering, North Carolina State University, Raleigh, North Carolina 27695, USA}
\affiliation{Department of Mathematics, North Carolina State University, Raleigh, North Carolina 27695, USA}

\author{Kausthubh Chandramouli}
\orcid{0009-0005-1676-1386}
\affiliation{Department of Electrical and Computer Engineering, North Carolina State University, Raleigh, North Carolina 27695, USA}

\author{Christian K\"ummerle}
\orcid{0000-0001-9267-5379}
\affiliation{School of Data, Mathematical, and Statistical Sciences, University of Central Florida, Orlando, Florida 32816, USA}

\author{Bojko N. Bakalov}
\orcid{0000-0003-4630-6120}
\affiliation{Department of Mathematics, North Carolina State University, Raleigh, North Carolina 27695, USA}

\author{Dror Baron}
\orcid{0000-0002-6371-8496}
\email{barondror@gmail.com}
\affiliation{Department of Electrical and Computer Engineering, North Carolina State University, Raleigh, North Carolina 27695, USA}

\date{\today}

\begin{abstract}
    Quantum state tomography (QST) is an indispensable tool for characterizing many-body quantum systems. However, due to the exponential scaling of the cost of the protocol with system size, many approaches have been developed for quantum states with specific structure, such as low-rank states. In this paper, we show how approximate message passing (AMP), an algorithmic framework for sparse signal recovery, can be used to perform low-rank QST. AMP provides asymptotically optimal performance guarantees for large sparse recovery problems, which suggests its utility for QST. We discuss the design challenges that come with applying AMP to QST, and show that by properly designing the AMP algorithm, we can reduce the reconstruction error by over an order of magnitude compared to existing approaches to low-rank QST. We also performed tomographic experiments on IBM Kingston and considered the effect of device noise on the reliability of the predicted fidelity of state preparation. Our work advances the state of low-rank QST and may be applicable to other quantum tomography protocols.
\end{abstract}

\maketitle

\section{Introduction}
A fundamental task in quantum computation is the ability to prepare a target quantum state. This task is accompanied by the need for a way to verify that the prepared state is indeed what was expected. There are many such techniques, including direct fidelity estimation~\cite{flammia_direct_2011}, certification~\cite{badescu_quantum_2017}, and classical shadows~\cite{huang_predicting_2020}. However, the most extensive characterization technique, which provides a full representation of the prepared quantum state, is \emph{quantum state tomography} (QST)~\cite{leonhardt_quantum-state_1995,leibfried_experimental_1996,paris_quantum_2004}. QST is the process of estimating the full state of a quantum system from noisy measurement data. QST can even serve as a diagnostic tool: the reconstructed state can reveal errors in the quantum device, such as coherent gate overrotations,  which can facilitate improvements in the performance of a quantum computer. For this reason, QST has been referred to as the ``gold standard"~\cite{stricker_experimental_2022} for quantum state characterization.

Although QST is a comprehensive protocol, its utility is limited by experimental and computational costs that scale exponentially with system size. There are two main challenges that arise from this scaling. The first is to collect enough measurement data to reconstruct a quantum state with an exponential number of degrees of freedom. For a gate-based quantum computer, the measurement data are obtained by executing a set of circuits whose empirical outcome distributions can be used to estimate the expectation values of a set of quantum observables~\cite{smithey_measurement_1993}. Although the number of circuits required to collect these measurements can be multiple orders of magnitude less than the number of observables, the number of circuits nevertheless scales exponentially with system size.

The second challenge is to store and process the measurement information on a classical computer to perform the reconstruction. One must store the sensing map defined by the observables, the experimentally collected measurement data, and the estimated quantum state. At the scale of utility-regime quantum algorithms, which are likely to require hundreds or thousands of qubits~\cite{preskill_quantum_2018}, QST quickly becomes impractical.

In order to improve the resource scaling for the QST protocol, one can impose structure on the target state to be reconstructed, such as low-rank structure~\cite{gross_quantum_2010, gross_recovering_2011, guta_rank-based_2012, kim_fast_2023}. A low-rank approximation is natural, as the rank of the density matrix corresponds to the number of pure states in the underlying mixed state ensemble. With this approximation, it is possible to apply \emph{compressed sensing} techniques such as constrained trace minimization~\cite{gross_quantum_2010} or the matrix LASSO~\cite{flammia_quantum_2012} to both reduce the number of measurements that must be collected---i.e., the number of circuits that must be executed---and improve the computational resource requirements for performing QST.

A powerful compressed sensing technique that has thus far remained unexplored in the QST setting is \emph{approximate message passing} (AMP)~\cite{DMM2009, bayati_dynamics_2011, feng_unifying_2022}. AMP refers to a class of iterative algorithms that can be used to solve linear inverse problems. These algorithms are particularly useful for two reasons. First, they are easily tailored to incorporate prior information about the structure of the signal, such as low-rank structure~\cite{matsushita_low-rank_2013, deshpande_information-theoretically_2014, deshpande_asymptotic_2016, montanari_non-negative_2014, kabashima_phase_2016, lesieur_constrained_2017, fletcher_iterative_2018, berthier_state_2020, montanari_estimation_2021}. Second, provided that certain technical conditions on the sensing map and signal are met, AMP achieves asymptotically optimal reconstruction error in the \emph{large-system limit}~\cite{bayati_lasso_2012,donoho_information-theoretically_2012, sur_likelihood_2019}, where the ratio of the number of measurements to the dimension of the signal tends to a constant. These characteristics make AMP a promising candidate for low-rank QST, where the quantum state has a known structure, and the dimension of the quantum state grows exponentially. In this paper, we explore the application of AMP to low-rank QST. Our results demonstrate that an AMP-based approach to low-rank QST can reduce the reconstruction error by {\em over an order of magnitude} compared to existing low-rank QST techniques (see Figures~\ref{fig:observable-comparison} and \ref{fig:shot-comparison}).

The paper is organized as follows. Section~\ref{sec:preliminaries} introduces the necessary notation and mathematical background. Section~\ref{sec:qst} provides the reader with an overview of the QST problem. Section~\ref{sec:amp} describes AMP and the design challenges that come with applying AMP to QST. Specifically, Section~\ref{subsec:AMP_QST} shows how to overcome these challenges by appropriately designing the AMP algorithm. Section~\ref{sec:numerical-results} contains our simulated numerical results, which demonstrate the advantage of our AMP approach compared to two existing low-rank QST techniques---maximum likelihood estimation (MLE)~\cite{teo_incomplete_2012,teo_quantum-state_2011,teo_numerical_2013,sacchi_maximum-likelihood_2001,rehacek_diluted_2007,lvovsky_iterative_2004,jezek_quantum_2003,hradil_quantum-state_1997,goncalves_local_2012,glancy_gradient-based_2012,fiurasek_maximum-likelihood_2001,blume-kohout_hedged_2010,baumgratz_scalable_2013-1,aditi_rigorous_2025} and momentum-inspired factored gradient descent (MiFGD)~\cite{kim_fast_2023}---across a range of quantum states. Section~\ref{sec:experimental-results} discusses the necessary considerations for deploying our AMP algorithm to characterize quantum states prepared on real quantum devices. In particular, Section~\ref{subsec:settings-comp} shows how to reduce the runtime on a quantum computer needed for QST by creating circuits based on measurement settings instead of observables, and Section~\ref{subsec:pred-fid-under-noise} uses Qiskit Aer~\cite{javadi-abhari_quantum_2024} to investigate the effects of device noise on QST reconstruction quality. Section~\ref{subsec:ibmq} provides a demonstration of experimental QST on IBM Kingston, which is informed by these considerations. We conclude the paper with a brief summary and discussion in Section~\ref{sec:disc}. Appendix~\ref{appen:photon-error-model} contains a discussion of an error model relevant to photonic devices, which serves as motivation for Section~\ref{subsec:higher-rank-states}. In Appendix~\ref{appen:meas-to-obs}, we provide additional details on the measurement settings discussed in Section~\ref{subsec:settings-comp}.

\section{Preliminaries} \label{sec:preliminaries}

{\bf Notation for vectors and matrices.}
Our notation resembles that of~\cite{nielsen_quantum_2012}. The complex conjugate of a complex number $w$ is denoted by $\bar{w}$, and its modulus by $|w|$. We refer to the number of qubits as $n$, and denote the Hilbert space by $\mathcal{H} = (\C^2)^{\otimes n} \cong \C^d$, where $d = 2^n$ and $\otimes$ is the Kronecker product. The transpose of a matrix $\rho$ is given by $\rho^T$, and its Hermitian conjugate by $\rho^\dagger = \bar{\rho}^T$. In $\C^2$, we have the computational basis vectors
\begin{equation}
    | 0 \rangle =
    \begin{pmatrix}
        1 \\
        0
    \end{pmatrix},
    \quad
    |1\rangle = \begin{pmatrix}
        0 \\
        1
    \end{pmatrix}.
\end{equation}
The Hermitian conjugate $| \psi \rangle^\dagger$ of a vector $\ket\psi\in\mathcal H$ is denoted by $\langle \psi |$.

Let $\H^{d \times d}$ be the real vector space of $d \times d$ complex Hermitian matrices---i.e., those matrices that can be expressed as a linear combination of Hermitian matrices with real coefficients---and $\mathcal{S}(\mathcal{H}) \subset \H^{d \times d}$ be the set of $d \times d$ density matrices. Recall that a \emph{density matrix}~\cite{nielsen_quantum_2012} is a Hermitian matrix $\rho$ with $\Tr \rho = 1$ that is positive semidefinite (PSD). The PSD condition, also denoted $\rho \succeq 0$, means that $\bra\psi\rho\ket\psi \ge 0$ for all $\ket\psi\in\mathcal H$.
Any Hermitian matrix $\rho$ has a spectral decomposition given by
\begin{equation}
    \rho = \sum_{k} \lambda_k \ket{\psi_k} \bra{\psi_k},
\end{equation}
where $\lambda_k \in \R$ are the eigenvalues (real because $\rho$ is Hermitian) and $\ket{\psi_k} \in \mathcal H$ are the eigenvectors of $\rho$. The rank of $\rho$, denoted by $\rank \rho$, is equal to the number of nonzero eigenvalues. The PSD property $\rho \succeq 0$ is equivalent to $\lambda_k\ge0$ for all $k$. In this case, we define the square root of $\rho$ by
\begin{equation}
    \sqrt{\rho} = \sum_{k} \sqrt{\lambda_k} \ket{\psi_k} \bra{\psi_k}.
\end{equation}
A pure state is defined as a density matrix $\rho$ of rank $1$; equivalently, $\rho=\ket\psi\bra\psi$, where the state vector $\ket\psi\in\mathcal H$ has $\braket{\psi}=1$ and is determined up to a phase, i.e., up to multiplication by $e^{i\theta}$ for $\theta\in\mathbb R$.

We will denote the $k$-th entry of a column vector $v$ by $v_k$. For a matrix $A$, we let $A_k$ be the $k$-th row of $A$, and $A_{k,l}$ the entry of $A$ at row $k$ and column $l$. For a linear transformation $\mathcal{A}\colon \C^a \rightarrow \C^b$, its matrix representation in a chosen basis is denoted by $\mathcal{M}(\mathcal{A}) \in \C^{b \times a}$. We will use $\mathcal{A}$ when referring to this transformation abstractly, and $\mathcal{M}(\mathcal{A})$ when the matrix realization is more pertinent, e.g., in storage on a classical computer.

We also make use of the vectorization map $\vectorize\colon \C^{d \times d} \rightarrow \C^{d^2}$ defined for $X \in \C^{d \times d}$ as $x = \vectorize(X)$ with
\begin{equation}
    x_{(k-1)d + l} = X_{k,l} \;\; (1 \leq k \leq d,\;1\leq l \leq d).
\end{equation}
We also consider the inverse map $\vectorize^{-1}\colon \C^{d^2} \rightarrow \C^{d \times d}$. Viewing $X$ as $\vectorize^{-1}(x)$ simply flips the previous equation, so that $X_{k,l} = x_{(k-1)d+l}$. 

The Pauli matrices $I$, $X$, $Y$, and $Z$ are given by
\begin{align}\label{pauli-1}
    I &=
    \begin{pmatrix}
        1 & 0 \\
        0 & 1
    \end{pmatrix}, &
    X &=
    \begin{pmatrix}
        0 & 1 \\
        1 & 0
    \end{pmatrix}, \\ \label{pauli-2}
    Y &=
    \begin{pmatrix}
        0 & -i \\
        i & 0
    \end{pmatrix}, &
    Z &=
    \begin{pmatrix}
        1 & 0 \\
        0 & -1
    \end{pmatrix}.
\end{align}
The set of all $n$-qubit Pauli strings $\{I, X, Y, Z\}^{\otimes n}$ is denoted by $\mathcal{P}_n$. We refer to each tensor factor in a Pauli string $P \in \mathcal{P}_n$ as a letter. The $d \times d$ identity map is denoted by $\mathbb{I}_d$ (and therefore $I = \mathbb{I}_2$).

{\bf Reconstruction error metrics.}
In this paper, we discuss two quality metrics for the QST reconstruction. The first is the normalized mean squared error (NMSE). For density matrices $\rho, \varsigma \in \mathcal{S}(\mathcal{H})$, where $\varsigma$ is an estimate for $\rho$, the NMSE is given by
\begin{equation} \label{eq:def-nmse}
    \NMSE(\rho, \varsigma) = \frac{\norm{\varsigma - \rho}_F^2}{\norm{\rho}_F^2},
\end{equation}
where $\norm{\rho}_F = \sqrt{\Tr [\rho \rho^\dagger]} = \sqrt{\sum_{k,l}|\rho_{k,l}|^2}$ is the Frobenius norm of $\rho$. 

We also consider the state fidelity $F(\rho, \varsigma)$ given by
\begin{equation} \label{eq:def-state-fid}
    F(\rho, \varsigma) = \left( \Tr  \sqrt{ \sqrt{\rho} \varsigma \sqrt{\rho}} \right)^2.
\end{equation}
The state fidelity $F(\rho, \varsigma) \in [0,1]$ can be interpreted, as detailed in Section 9.2 of \cite{wilde2013quantum}, as the probability that the state $\rho$ would pass as identical to the state $\varsigma$ to an observer who knows $\varsigma$, if both density matrices represent pure states. We define the state infidelity as $1 - F(\rho, \varsigma)$. If $\varsigma \notin \mathcal{S}(\mathcal{H})$, then we can define $F(\rho, \varsigma)$ by first computing the projection of $\varsigma$ onto $\mathcal{S}(\mathcal{H})$. 
This projection is defined as follows:
\begin{equation} \label{eq:def-proj}
    \proj_{\mathcal{S}(\mathcal{H})}(\varsigma) = \frac{\displaystyle\sum_{k:\mu_k > 0} \mu_k \ket{\varphi_k} \bra{\varphi_k}}{\displaystyle\sum_{k:\mu_k > 0} \mu_k} \,,
\end{equation}
where $\varsigma = \sum_k \mu_k \ket{\varphi_k} \bra{\varphi_k}$ has eigenvalues $\mu_k\in\mathbb R$ (real because $\varsigma$ is Hermitian) and eigenvectors $\ket{\varphi_k}\in\mathcal H$. We note that this projection is not optimal with respect to the Frobenius norm---see, e.g., references~\cite{wang_projection_2013,duchi_efficient_2008,shalev-shwartz_efficient_2006,chen_projection_2011} for an algorithm that performs optimal Euclidean-norm projection of the eigenvalues onto the probability simplex and is thus optimal with respect to the Frobenius norm.

The $L^2$ norm of a vector $v\in\mathcal H$ is given by
\begin{equation}
    \norm{v}_2 =\sqrt{\braket{v}} = \left( \sum_{k=1}^d |v_k|^2 \right)^{1/2},
\end{equation}
where $v_k\in\mathbb C$ are the coordinates of $v$.

{\bf Specific quantum states.} In this paper, we apply QST to GHZ, Hadamard, and W states, whose state vectors are given by~\cite{nielsen_quantum_2012}:
\begin{align}
        |\text{GHZ}(n)\rangle &= \frac{| 0 \rangle^{\otimes n} + | 1 \rangle^{\otimes n}}{\sqrt{2}}, \\
        |\text{Hadamard}(n)\rangle &= \left( \frac{ |0 \rangle + |1 \rangle}{\sqrt{2}} \right)^{\otimes n}, \\
        |\text{W}(n)\rangle &= \frac{1}{\sqrt{n}} \sum_{i=0}^{n-1} |0\rangle^{\otimes i} \otimes | 1 \rangle \otimes | 0 \rangle^{\otimes (n- i -1)}.
\end{align}
The corresponding density matrices are defined as:
\begin{align}
\rho_{\text{GHZ}(n)} &= \ket{\text{GHZ}(n)} \bra{\text{GHZ}(n)}, \\
\rho_{\text{Hadamard}(n)} &= \ket{\text{Hadamard}(n)} \bra{\text{Hadamard}(n)}, \\
\rho_{\text{W}(n)} &= \ket{\text{W}(n)} \bra{\text{W}(n)}.
\end{align}

We also consider random states $\rho_{\text{Random}(n, r)}$ of rank $r$. 
To generate such a state, we generate $r$ random vectors $|\psi_k\rangle\in \mathcal{H}$ satisfying $| \psi_k \rangle _j \sim \mathcal{CN}(0,1)$, and a random $p \in \R^r$ satisfying $p_k \sim \mathcal{U}(0,1)$,
where $\mathcal{CN}(0, \sigma^2)$ denotes the complex normal distribution, and $\mathcal{U}(0,1)$ is the uniform distribution over the interval $(0,1)$.
From $\ket{\psi_1}, \dots, \ket{\psi_r}$ and $p$, we compute $| \widetilde{\psi}_k \rangle = | \psi_k \rangle / \norm{ | \psi_k \rangle}_2$ and $\widetilde{p} =  p / \sum_k p_k$, so that $\sum_k \widetilde{p}_k = 1$ and $\widetilde{p}$ can be interpreted as a probability vector. Then a state $\rho_{\text{Random}(n, r)}$ is given by
\begin{equation}
\rho_{\text{Random}(n, r)} = \sum_{k=1}^r \widetilde{p}_k |\widetilde{\psi}_k \rangle \langle \widetilde{\psi}_k |.
\end{equation}

\section{Quantum State Tomography} \label{sec:qst}
\textbf{Formulation.} QST~\cite{leonhardt_quantum-state_1995,leibfried_experimental_1996,paris_quantum_2004} addresses the task of reconstructing a particular $n$-qubit quantum state $\rho^* \in \mathcal{S}(\mathcal{H})$ using noisy measurement data. Each measurement corresponds to a quantum observable whose expectation value we estimate by repeatedly preparing the quantum state and measuring the observable, and then taking the sample average of the measured eigenvalues. 

In this work, we will restrict ourselves to the set $\mathcal{P}_n$ of Pauli observables, which forms an orthonormal basis for the real vector space of Hermitian matrices $\H^{d \times d}$. Any $P_k \in \mathcal{P}_n$ has eigenvalues $\pm 1$. If we measure $P_k$ for a total of $N$ shots and obtain the $+1$ eigenvalue $N_k$ times, then we must measure the $-1$ eigenvalue $N - N_k$ times, and therefore the sample mean $y_k$ is
\begin{align}
    y_k &= \frac{1}{N} \bigl(N_k (+1) + (N - N_k) (-1) \bigr) \notag \\
    &= \frac{1}{N} \left(2 N_k - N \right) \label{eq:fk} \\
    &= 2f_k - 1, \notag
\end{align}
where $f_k = N_k / N$. The sample mean $y_k$ is equal to the true expectation value $\Tr [P_k \rho^*]$ plus binomial shot noise $z_k$. As we increase the number of shots $N$, the noise $z_k$ will tend to decrease in magnitude. 

The quantum state $\rho^*$ has possibly $\mathcal{O}(d^2)$ degrees of freedom, where $\mathcal{O}(\cdot)$ refers to the typical Big $\mathcal{O}$ notation \cite{cormen2022introduction}. Hence, we sample $M$ such Pauli observables $P_k$ from $\mathcal{P}_n$ and assemble the sample means $y_k$ into a single data vector $y \in \R^M$. The $M$ Pauli observables collectively define a sensing map $\mathcal{A}\colon \H^{d \times d} \rightarrow \R^M$, whose action on $\rho^*$ is given by
\begin{equation} \label{eq:def-a}
    \mathcal{A}(\rho^*)_k = \Tr [P_k \rho^*], \quad 1\le k\le M.
\end{equation}
The vector $\mathcal{A}(\rho^*)$ contains the exact expectation values for each observable. Since each sample mean $y_k$ is equal to the true expectation value $\Tr [P_k \rho^*]$ plus binomial shot noise $z_k$, we have
\begin{equation} \label{eq:qst-formulation}
    y = \mathcal{A}(\rho^*)+z.
\end{equation}
The task of any QST algorithm is to generate an estimate $\widehat{\rho} \in \mathcal{S}(\mathcal{H})$ for $\rho^*$ using~\eqref{eq:qst-formulation}.

The action of $\mathcal{A}$ on $\rho^*$ can also be expressed as a matrix-vector product. To see how, we note that $P_k$ is Hermitian, and write~\eqref{eq:def-a} as:
\begin{align*}
    \mathcal{A}(\rho^*)_k = \sum_{i=1}^d (P_k \rho^*)_{i,i} &= \sum_{i,j=1}^d (P_k)_{i,j} \rho^*_{j,i} \\
    &= \sum_{i,j=1}^d (\bar{P}_k)_{j,i} \rho^*_{j,i}.
\end{align*}
The sum is over the products of the corresponding entries of $\bar{P}_k$ and $\rho^*$, and therefore:
\begin{align*}
    \mathcal{A}(\rho^*)_k &= (\vectorize \bar{P}_k)^T (\vectorize \rho^*) \\
    &= (\vectorize P_k)^\dagger (\vectorize \rho^*).
\end{align*}
We obtain
\begin{align}
    \mathcal{A}(\rho^*) &= \mathcal{M}(\mathcal{A}) (\vectorize \rho^*), \\
    \mathcal{M}(\mathcal{A})_k &= (\vectorize P_k)^\dagger. \label{eq:def-m-of-a}
\end{align}
The adjoint map $\mathcal{A}^\dagger$ is given by
\begin{equation} \label{eq:def-a-dagger}
    \mathcal{A}^\dagger(y) = \sum_{k=1}^M y_k P_k,
\end{equation}
and in matrix form,
\begin{equation}
    \mathcal{A}^\dagger(y) = \vectorize^{-1}\bigl( \mathcal{M}(\mathcal{A})^\dagger y \bigr).
\end{equation}

\textbf{Prior art.} The space of QST algorithms is vast, including linear inversion~\cite{qi_adaptive_2017,hou_full_2016,qi_quantum_2013,smolin_efficient_2012,james_measurement_2001,dariano_detection_1994}, maximum entropy~\cite{buzek_quantum_2000,goncalves_quantum_2013,gupta_maximal_2021,teo_incomplete_2012,teo_quantum-state_2011}, Bayesian tomography~\cite{tanaka_bayesian_2005,schack_quantum_2001,rau_evidence_2010,huszar_adaptive_2012,granade_practical_2016,fuchs_finetti_2004,buzek_reconstruction_1998,blume-kohout_accurate_2006,blume-kohout_optimal_2010,audenaert_quantum_2009}, maximum likelihood estimation (MLE)~\cite{teo_incomplete_2012,teo_quantum-state_2011,teo_numerical_2013,sacchi_maximum-likelihood_2001,rehacek_diluted_2007,lvovsky_iterative_2004,jezek_quantum_2003,hradil_quantum-state_1997,goncalves_local_2012,glancy_gradient-based_2012,fiurasek_maximum-likelihood_2001,blume-kohout_hedged_2010,baumgratz_scalable_2013-1,aditi_rigorous_2025}, gradient descent~\cite{wang_efficient_2024,teo_numerical_2013,shang_superfast_2017,rehacek_diluted_2007,kyrillidis_provable_2018,kim_local_2023,kim_fast_2023,hsu_quantum_2024,goncalves_projected_2016,glancy_gradient-based_2012,gaikwad_gradient-descent_2025,ferrie_self-guided_2014,bolduc_projected_2017}, neural networks~\cite{wei_neural-shadow_2024,torlai_machine-learning_2020,torlai_neural-network_2018,tiunov_experimental_2020,smith_efficient_2021,schmale_efficient_2022,rocchetto_experimental_2019,quek_adaptive_2021,palmieri_experimental_2020,lohani_machine_2020,lloyd_quantum_2018,liu_variational_2020,kuzmin_learning_2024,koutny_neural-network_2022,glasser_neural-network_2018,cha_attention-based_2022,carrasquilla_how_2021,carrasquilla_reconstructing_2019,carleo_constructing_2018,ahmed_quantum_2021,ahmed_classification_2021}, projected classical shadows~\cite{qin_enhancing_2025}, and compressed sensing~\cite{yang_effective_2017,tonolini_reconstructing_2014,steffens_experimentally_2017,shabani_efficient_2011,riofrio_experimental_2017,liu_experimental_2012,kyrillidis_provable_2018,kueng_low_2017,kosut_quantum_2008,gross_quantum_2010,flammia_quantum_2012,ahn_adaptive_2019}. These algorithms can be broadly divided into two classes: full quantum state tomography (FQST) methods and compressed quantum state tomography (CQST) methods. 

For FQST methods, $\mathcal{A}$ is full rank, i.e., we use all $d^2$ Pauli observables. However, since $d$ grows exponentially in $n$, FQST quickly becomes impractical. Both the experimental cost of collecting data for $d^2$ observables and the storage of $\mathcal{M}(\mathcal{A})$ and $\widehat{\rho}$ pose a challenge for running FQST on even a dozen qubits.

CQST methods reduce the number of observables required to perform QST by imposing additional structure on $\rho^*$. This structure reduces the amount of information necessary to perform QST. Prior work has explored permutationally invariant states~\cite{toth_permutationally_2010,schwemmer_experimental_2014,moroder_permutationally_2012}, matrix-product states~\cite{qin_quantum_2024,lidiak_quantum_2022,lanyon_efficient_2017,han_density_2022,cramer_efficient_2010,baumgratz_scalable_2013}, and, of particular interest to us in this paper, low-rank quantum states~\cite{odonnell_efficient_2016,liu_universal_2011,guta_rank-based_2012,baldwin_strictly-complete_2016}. A low-rank state $\rho^*$ can be expressed as $\rho^* = UU^\dagger$, where $U \in \C^{d \times r}$ and $r \ll d$. In this case, $\rho^*$ only has $(2d - r)r = \mathcal{O}(rd)$ degrees of freedom. It has been shown~\cite{gross_quantum_2010} that $M=\mathcal{O}(rdn^2)$ Pauli observables suffice to recover any state $\rho^*$, such that $\rank \rho^* \leq r$, with high probability. Although the number of observables required is still exponential in $n$, it is an improvement over $d^2$.

Many interesting states are low-rank, including pure states (which have rank equal to one). In Section~\ref{subsec:higher-rank-states} and Appendix~\ref{appen:photon-error-model}, we describe a noise model on a photonic quantum device for which the rank grows at most linearly in $n$, which also generates low-rank states for sufficiently large $n$.

As we detail further in the next section, AMP is an iterative technique that, with the appropriate design modifications, can be applied to QST for low-rank states.

\section{Approximate Message Passing} \label{sec:amp}

AMP~\cite{DMM2009, bayati_dynamics_2011, feng_unifying_2022} refers to a class of iterative algorithms that solve linear inverse problems. This section lays out the details of our approach to QST using AMP. Because our presentation is somewhat involved, we begin with an overview in Section~\ref{subsec:amp-overview}. The details appear in Sections \ref{subsec:initial_AMP} and \ref{subsec:AMP_QST}, once the big picture has been laid out.

\subsection{Overview} \label{subsec:amp-overview}

{\bf Standard AMP.}
The presentation commences in Section~\ref{subsec:initial_AMP}, where we describe a standard version of AMP. While the QST
density matrix recovery problem is somewhat mismatched with the ideal setting studied in the AMP literature (details below),
standard AMP has been studied extensively in the literature, which will allow us to provide insights and theoretical properties.
Importantly, 
under the appropriate technical conditions~\cite{bayati_dynamics_2011},
AMP achieves asymptotically optimal mean squared error (MSE) performance equivalent to the Bayes-optimal estimator. Based on standard AMP, Section~\ref{subsec:initial_AMP} concludes with an initial AMP-based algorithm for QST.

\begin{figure}
    \centering
    \includegraphics[width=\linewidth]{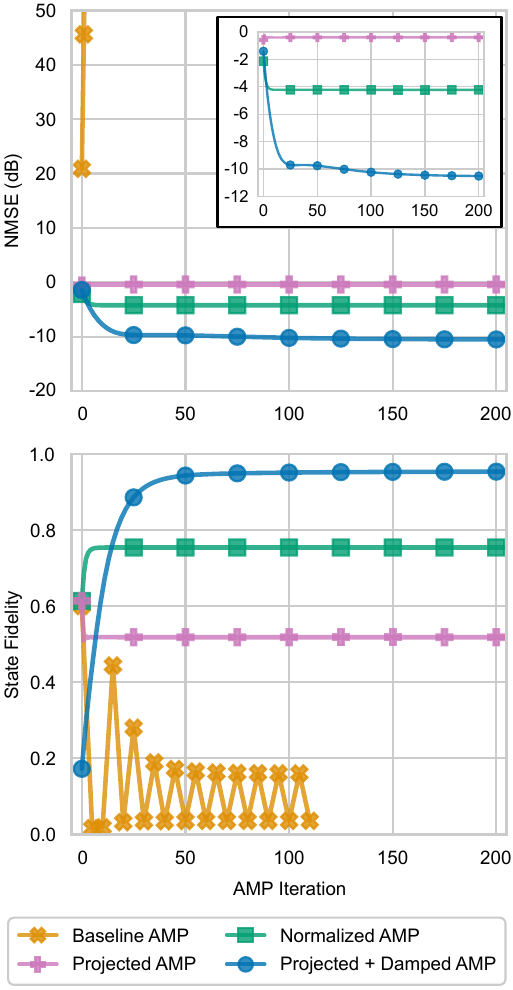}
    \caption{{A comparison of different AMP approaches for QST}. We reconstruct a rank-3 5-qubit random state ($M=384$ observables, $N=1024$ shots per observable)
    using variations of the AMP algorithm and show the reconstruction quality as measured by the normalized mean squared error (NMSE) and state fidelity. The baseline AMP approach (orange crosses) diverges. After replacing the sensing operator $\mathcal{A}$~\eqref{eq:amp-alg-pseudodata} with the normalized version $\widetilde{\mathcal{A}}$~\eqref{eq:def-normalized-a} (green squares), AMP converges and produces an estimator $\widehat{\rho}$ for the true density matrix $\rho^*$. To improve the estimator, we incorporate the physical constraints of $\rho^*$, by using the projected singular value thresholding ($\PSVT$) denoiser~\eqref{eq:def-psvt}. Without damping (purple pluses), the $\PSVT$-based AMP algorithm does not recover $\rho^*$. With damping~\eqref{eq:damping} (blue circles), the $\PSVT$-based denoiser recovers $\rho^*$ with lower NMSE and higher state fidelity than the $\SVT$-based approach. The inset in the NMSE plot is a magnification of the normalized, projected, and projection-plus-damping results.}
    
    \label{fig:amp-design-comparison}
\end{figure}

{\bf Challenges for AMP in QST.}
Again, our initial AMP approach is somewhat mismatched with the QST problem setting (details in Section~\ref{subsec:AMP_QST}). First, considering the AMP measurement matrix, $B$, the matrix $B^{\dagger}B$ must resemble an identity matrix, which is not true for the QST sensing matrix $\mathcal{M}(\mathcal{A})$ given by Eq.~\eqref{eq:def-m-of-a}. Second, many of the theoretical results that have been derived for AMP require $B$ to have specific properties, which are explicitly stated in Section~\ref{subsec:initial_AMP}; $\mathcal{M}(\mathcal{A})$ does not have these properties. Third, the output of our reconstruction algorithm should be a physically valid density matrix $\widehat{\rho}$, such that $\widehat{\rho} \succeq 0$ and $\Tr \widehat{\rho} = 1$. However, a standard low-rank AMP denoiser based on singular value thresholding (SVT) does not guarantee that the matrix it returns is PSD or has unit trace.

{\bf AMP design.}
Section~\ref{subsec:AMP_QST} addresses these challenges by adapting the initial AMP approach of Section~\ref{subsec:initial_AMP} for QST. First, to address the incorrect normalization of $\mathcal{A}$, we define a normalized sensing map $\widetilde{\mathcal{A}}$. Second, we use a modified denoiser that forces each iteration in our estimation of $\rho^*$ to produce a PSD matrix with a unit trace, which ensures $\widehat{\rho} \in \mathcal{S}(\mathcal{H})$. In order to benefit from this modification, we apply \emph{damping}~\cite{vila_adaptive_2015}, which controls the assertiveness of the AMP algorithm. Although less is known about the theoretical properties of damped AMP, it has been previously applied successfully in other situations where the technical conditions of standard AMP are not met~\cite{vila_adaptive_2015}.

{\bf Numerical example.}
We conclude our overview with Figure~\ref{fig:amp-design-comparison}, where we show how our design choices enable density matrix reconstruction. We take $\rho^*$ to be a random rank-3 5-qubit state and set $M = 384$ ($=0.375d^2$) observables, $N = 1024$ shots per observable, and chart the difference between the estimate $\rho^t$ and $\rho^*$ at each iteration $t$ of the AMP algorithm. We plot both $\NMSE(\rho^*, \rho^t)$ and $F(\rho^*, \rho^t).$\footnote{For the standard AMP and normalized AMP approaches, the density matrix estimate $\rho^t$ at each iteration is not guaranteed to be physical, as the low-rank denoiser does not always produce a matrix with unit trace. Thus, we first project $\rho^t$ onto the set of density matrices~\eqref{eq:def-proj} before computing the state fidelity shown in Figure~\ref{fig:amp-design-comparison}.} The baseline AMP approach (orange crosses), based on~\cite{berthier_state_2020}, diverges since the QST sensing map $\mathcal{A}$ is not normalized. After replacing $\mathcal{A}$ with the normalized map $\widetilde{\mathcal{A}}$~\eqref{eq:def-normalized-a} (green squares), AMP converges and produces an estimator $\widehat{\rho}$ for the density matrix $\rho^*$. To improve the fidelity of this estimator, we incorporate the physical constraints of $\rho^*$, by using the projected singular value thresholding ($\PSVT$) denoiser~\eqref{eq:def-psvt}. By itself (purple pluses), the $\PSVT$-based AMP algorithm does not recover $\rho^*$. However, by incorporating damping~\eqref{eq:damping} (blue circles), the $\PSVT$-based AMP algorithm recovers $\rho^*$ with lower NMSE and higher state fidelity than the $\SVT$-based approach.

\subsection{Standard AMP and initial approach}
\label{subsec:initial_AMP}
{\bf Standard AMP.}
Consider the task of recovering an
unknown complex vector $x \in \C^N$ from $M$ linear measurements in the presence of additive 
white Gaussian noise (AWGN) $z$ with variance $\sigma_z^2$.
This task can be written as a linear inverse problem,
\begin{equation}
    y = Bx + z,
\end{equation}
where $y \in \C^M$ are the noisy linear
measurements,
$B\in \C^{M \times N}$,
and the entries of the noise $z$ are 
given by $z_k \sim \mathcal{CN}(0, \sigma_z^2)$.

To recover $x$, we apply AMP iterations comprised of the following equations
(\ref{eq:amp-alg-residual})--(\ref{eq:denoise_vt}), 
where $0 \leq t < t_{\text{max}}$ is the iteration index, and $x^t$ are estimates of the unknown $x$ 
in iteration~$t$:
\begin{align}
    r^t &= y - Bx^t + c_t r^{t-1}, \label{eq:amp-alg-residual} \\
    v^t &= x^{t} + B^\dagger r^t, \label{eq:amp-alg-pseudodata} \\
    x^{t+1} &= f_t\bigl( v^t \bigr),\label{eq:denoise_vt}
\end{align}
with initialization $x^0 = 0$ and $r^{-1} = 0$. The coefficients $c_t$ are given by
\begin{equation}\label{eq:ct}
    c_t = \frac{1}{M} \divword f_t (v^{t-1}) = \frac{1}{M} \sum_{k=1}^N \frac{\partial f_t(v^{t-1})_k}{\partial v^{t-1}_k},
\end{equation}
$r^t \in \C^{M}$ is the \emph{residual} or unexplained part of the measurements,
$v^t$ is the {\em pseudo-data},
and 
$f_t\colon \C^N \rightarrow \C^{N}$ is a {\em denoiser} function that
incorporates the known prior information about the structure of $x$ to generate an estimate from the pseudo-data $v^t$. 
We note that we use $B^\dagger$ instead of $B^T$ as in real-valued AMP~\cite{DMM2009}, because $x$ has complex entries~\cite{maleki_asymptotic_2013}. 

In (\ref{eq:amp-alg-residual}),
the \emph{Onsager reaction term} $c_t r^{t-1}$
ensures that the estimation error is uncorrelated with the ground truth signal $x$ and is essential for obtaining accurate signal recovery; see~\cite{Thoules1977, DMM2009}. The denoiser function $f_t$ relies on statistical information about $x$ and the noise within the pseudo-data $v^t$. A common choice of denoiser function is conditional expectation,
$f_t(v^t)=\mathbb{E}[X|V^t=v^t]$, although other denoiser functions are also often used.

{\bf Properties of standard AMP.}
We provide a technical condition for AMP known as the {\em large system limit}.

\smallskip
\textbf{Condition~1.} For $M, N \rightarrow \infty$ with $M/N$ fixed and $1 \leq m \leq M$, $1 \leq n \leq N$, the entries $B_{m,n}$ are independent and identically distributed (i.i.d.) samples from $\mathcal{CN}(0,1/M)$.
Moreover,
\begin{equation} \label{eq:mat-expected-value-1}
    \mathbb{E}\bigl[ B^\dagger B \bigr] = \mathbb{I}_N.
\end{equation}

Under Condition~1, it is known~\cite{bayati_dynamics_2011} that the sequence of AMP estimates achieves asymptotically optimal MSE performance equivalent to the Bayes-optimal estimator. 
Additionally, Condition~1 ensures that
\begin{equation}
    v^t_{k} \stackrel{d}{=} x_k + \mathcal{CN}(0, \sigma_t^2), \quad 1\le k\le N,
\end{equation}
where $\stackrel{d}{=}$ means equal in distribution and $\sigma_t^2$ is the noise variance at iteration $t$. 
This equation is why we refer to $v^t$ as the pseudo-data: it is equal in distribution to the ground truth $x$ plus AWGN. The statistical structure of the pseudo-data implies that $\sigma_t^2$ evolves according to a {\em state evolution} formalism~\cite{DMM2009, Montanari2012}. Further, $\sigma_t^2$ 
satisfies~\cite{metzler_denoising_2016}
\begin{equation}\label{eq:sigmat}
    \sigma_t^2 \approx \frac{\norm{r^t}^2_2}{M}.
\end{equation}
These asymptotic performance guarantees under Condition~1 suggest that AMP may be well suited for QST, where the system size grows exponentially in the number of qubits.

{\bf Baseline AMP approach.}
In light of our discussion of standard AMP, we describe an initial AMP approach to QST, which is based on~\cite{berthier_state_2020}.
The unknown vector $x$ in standard AMP is
the unknown 
$d \times d$ density matrix $\rho^*$, and the measurement matrix $B$ is given by $\mathcal{M}(\mathcal{A})$~\eqref{eq:def-m-of-a}. Since $\rho^*$
is low-rank, an appropriate choice for the denoiser $f_t$ is the singular value thresholding (SVT) operator~\cite{Cai-2010Singular}. For a complex matrix $X$ 
with singular values $s_k$, left singular vectors $u_k$, and right singular vectors $v_k$, and a chosen threshold $\tau$, the SVT operator is given by
\begin{equation} \label{eq:svt-def}
    \SVT(X; \tau) = \sum_k (s_k - \tau)_+ u_k v_k^\dagger,
\end{equation}
where
\begin{equation}
    x_+ = 
    \begin{cases}
        x &\text{ if } x \geq 0 \\
        0 &\text{ if } x < 0.
    \end{cases}
\end{equation}
At each iteration of the AMP algorithm, the threshold $\tau_t$ can be chosen to be proportional to the noise level $\sigma_t$ and the size of $\rho^*$:
\begin{equation}
    \tau_t = \alpha  \sigma_t \sqrt{d}, \quad \sigma_t \approx \frac{\norm{r^t}_F}{\sqrt{M}},
\end{equation}
where $\alpha$ is a tunable proportionality constant; we set $\alpha = 2$~\cite{berthier_state_2020} throughout this paper. 

We compute the coefficients $c_t$ \eqref{eq:ct} using the Monte Carlo approach suggested in~\cite{ramani_monte-carlo_2008}. Fixing $\varepsilon$ small and taking $h \in \C^N$ with entries $h_{k} \sim \mathcal{CN}(0,1)$, we approximate $c_t$ by
\begin{equation} \label{eq:divergence-monte-carlo}
     \widehat{c_t} = \frac{1}{M}  \mathbb{E}_h \left[h^\dagger \cdot \left( \frac{f_t(v^{t-1} + \varepsilon h) - f_t(v^{t-1}) }{\varepsilon} \right) \right].
\end{equation}
The expectation $\mathbb{E}_h$ over $h$ can be approximated by taking $K$ such samples $h$ and averaging. For $d \gg 1$, it is often sufficient to set $K=1$, which we do in this paper. Our numerical experiments suggest that setting $K=10$ yields a negligible improvement in the estimation accuracy of $\widehat{c_t}$ compared to $K=1$, while increasing the runtime due to additional calls to the denoiser. Although a closed-form expression exists for $c_t$ when the SVT is employed as a denoiser~\cite{candes_unbiased_2013,donoho_minimax_2014}, we use~\eqref{eq:divergence-monte-carlo} instead because it provides a straightforward way to estimate $c_t$ when $f_t$ is a black-box denoiser.

\subsection{Design adaptations for QST}
\label{subsec:AMP_QST}

The QST problem setup poses several challenges to the framework outlined in the previous section. First, the noise $z$ is shot noise instead of AWGN. Moreover, the QST sensing matrix $\mathcal{M}(\mathcal{A})$ consists of vectorized Pauli matrices, whose entries are not i.i.d. Gaussian. A direct application of the algorithm outlined in the previous section to~\eqref{eq:qst-formulation} diverges, as is shown in Figure~\ref{fig:amp-design-comparison}. To address these issues, we make the following modifications.

{\bf Rescaling.} To address the divergence of the AMP algorithm, we normalize $\mathcal{A}$. Combining~\eqref{eq:def-a} and~\eqref{eq:def-a-dagger}, we have that:
\begin{equation}
    \mathcal{A}^\dagger \mathcal{A}(\rho^*) = \sum_{k=1}^M \Tr[P_k \rho^*] P_k.
\end{equation}
When we sum over all $d^{2M}$ ways to sample $M$ Paulis from $\mathcal{P}_n$ (the set of all $n$-qubit Pauli strings), with replacement, each of the $d^2$ unique Pauli strings is sampled $M  d^{2(M-1)}$ times. Thus, if we sample with replacement uniformly at random, then
\begin{align}
    \mathbb{E}\bigl[\mathcal{A}^\dagger \mathcal{A} (\rho^*) \bigr] &= \frac{M  d^{2(M-1)}}{d^{2M}} \sum_{k=1}^{d^2} \Tr[P_k \rho^*] P_k \\
    &= \frac{M}{d} \rho^*,
\end{align}
where we used the fact that (see, e.g., \cite{nielsen_quantum_2012})
\begin{equation}
\rho^* = \frac1d\sum_{k=1}^{d^2} \Tr[P_k \rho^*] P_k.
\end{equation}
Hence, $\mathbb{E}[\mathcal{A}^\dagger \mathcal{A}] = M\mathbb{I}_d/d$. In alignment with~\eqref{eq:mat-expected-value-1}, we want an operator $\widetilde{\mathcal{A}}$ that satisfies $\mathbb{E} [ \widetilde{\mathcal{A}}^\dagger \widetilde{\mathcal{A}} ] = \mathbb{I}_d$, and therefore we set
\begin{equation} \label{eq:def-normalized-a}
    \widetilde{\mathcal{A}} = \sqrt{\frac{d}{M}} \mathcal{A}.
\end{equation}
We likewise compute a rescaling $\widetilde{y}$ of the measurement vector $y$:
\begin{equation}
    \widetilde{y} = \sqrt{\frac{d}{M}} y.
\end{equation}
With this rescaling, the AMP converges, as shown in Figure~\ref{fig:amp-design-comparison}. 
Note that rescaling $y$ also rescales the noise $z$; however, this rescaling is taken into account when estimating the noise variance $\sigma_t^2$ in \eqref{eq:sigmat}, because $r^t$ is also rescaled according to Eq.~\eqref{eq:algo1} below.

{\bf Projection and damping.}
To leverage the fact that $\rho^*$ is a density matrix, we modify the $\SVT$ denoiser. In particular, we define a projected singular value thresholding operator $\PSVT(\cdot; \tau)$, which is the composition of the SVT operator with a projection onto $\mathcal{S}(\mathcal{H})$:
\begin{equation} \label{eq:def-psvt}
    \PSVT(\rho^t; \tau_t) = \proj_{\mathcal{S}(\mathcal{H})} \SVT(\rho^t; \tau_t).
\end{equation}
By itself, the $\PSVT$ operator does not enable estimation of $\rho^*$, as shown in Figure~\ref{fig:amp-design-comparison}. However, by incorporating \emph{damping}~\cite{vila_adaptive_2015}, we can estimate $\rho^*$ with substantially lower reconstruction error than with the $\SVT$-based algorithm. Damping controls the assertiveness with which the AMP algorithm proceeds by replacing the update rule \eqref{eq:denoise_vt} with
\begin{equation} \label{eq:damping}
    \rho^{t+1} = \lambda f_t(v^t) + (1-\lambda)\rho^t, \quad 0 < \lambda \leq 1.
\end{equation}
Damping has been previously employed to improve the convergence of AMP when Condition~1 is not met~\cite{vila_adaptive_2015}. We note that, based on our numerical experiments, applying damping to the $\SVT$-based AMP algorithm does not improve the recovery quality.

{\bf Algorithm.} Our AMP algorithm for low-rank QST consists of the following iterative steps:
\begin{align}
    r^t &= \widetilde{y} - \widetilde{\mathcal{A}} (\rho^t) + \widehat{c_t} r^{t-1}, \label{eq:algo1} \\
    v^t &= \rho^t + \widetilde{\mathcal{A}}^\dagger(r^t), \label{eq:algo2} \\
    \rho^{t+1} &= \lambda \PSVT(v^t; \tau_t) + (1-\lambda) \rho^t. \label{eq:algo3} 
\end{align}
We set $\rho^0 = \mathbb{I}_d/d$, $r^{-1} = 0$, $\tau_t = 2 \sigma_t \sqrt{d}$, $\lambda = 0.01$, and $t_{\text{max}} = 2000$. The values of $\lambda$ and $t_\text{max}$ were conservatively chosen to ensure the convergence of the AMP algorithm for all numerical experiments. We take $\widehat{\rho}$ to be the final iterate produced by the AMP algorithm, i.e., $\widehat{\rho} = \rho^{t_{\text{max}}}$.

\section{Numerical results} \label{sec:numerical-results}
In this section, we demonstrate the performance of our AMP-based QST algorithm \eqref{eq:algo1}--\eqref{eq:algo3} using numerical experiments. We benchmark the reconstruction quality against two existing low-rank QST approaches---maximum likelihood estimation (MLE)~\cite{teo_incomplete_2012,teo_quantum-state_2011,teo_numerical_2013,sacchi_maximum-likelihood_2001,rehacek_diluted_2007,lvovsky_iterative_2004,jezek_quantum_2003,hradil_quantum-state_1997,goncalves_local_2012,glancy_gradient-based_2012,fiurasek_maximum-likelihood_2001,blume-kohout_hedged_2010,baumgratz_scalable_2013-1,aditi_rigorous_2025} and momentum-inspired factored gradient descent (MiFGD)~\cite{kim_fast_2023}---and show reconstruction on states with up to $n=8$ qubits.

\subsection{Software methods}
We run all simulations in Python using a high-performance cluster. For a given number of observables $M$, we randomly sample $M$ observables from $\mathcal{P}_n$ (the set of all $n$-qubit Pauli strings) without replacement. We simulate shot data for each Pauli observable $P_k$ by calculating its exact expectation value and then sampling $y_k$ from the corresponding binomial distribution. That is,
\begin{equation}
    y_k \sim \frac{2}{N}\mathcal{B}(N, p_k) - 1, \quad p_k = \frac{1}{2} \bigl( \Tr[P_k \rho^*] + 1 \bigr),
\end{equation}
where we used Eq.~\eqref{eq:fk} and $\mathcal{B}(N, p_k)$ is the binomial distribution with parameters $N$ and $p_k$. We also consider the case where $N \rightarrow \infty$, i.e., ``infinite shots," which we implement by setting $y_k = 2p_k - 1$. (Recall that $y_k$ is equal to the true expectation value $\Tr [P_k \rho^*]$ plus shot noise, and thus in the limit of infinite shots, the shot noise vanishes.)

Next, we note a memory optimization that applies to the AMP and MLE methods. Naively, it takes $\mathcal{O}(M d^2)$ memory to store $\mathcal{M}(\widetilde{\mathcal{A}})$, where $\mathcal{O}(\cdot)$ refers to the typical Big $\mathcal{O}$ notation \cite{cormen2022introduction}, and each entry is a 16-byte complex double. For $n=10$ and $M = d$ (which is almost certainly too few measurements when $\mathcal{O}(rd n^2)$ are needed), storing $\mathcal{M}(\widetilde{\mathcal{A}})$ in this way would require approximately $Md^2 \times 16 = d^3 \times 16 = (2^{10})^3 \times 16 = 16$ GB of memory, which is outside the range of many commercial laptops. To reduce the memory required while retaining the ability to compute $\widetilde{\mathcal{A}}(\rho^t)$ and $\widetilde{\mathcal{A}}^\dagger (r^t)$ using fast matrix-vector multiplication, we factor $\mathcal{M}(\widetilde{\mathcal{A}})$ as follows. Let $n_k$ be the number of Pauli $Y$ matrices that occur in the Pauli string $P_k$. Then we write $\mathcal{M}(\widetilde{\mathcal{A}})$ as $DR$, where $D \in \C^{M \times M}$ is a diagonal matrix with entries
\begin{equation}\label{eq:D}
    D_{k,k} = \sqrt{\frac{d}{M}}i^{n_k}, \quad 1 \leq k \leq M,
\end{equation}
and $R \in \mathbb{Z}^{M \times d^2}$ is an integer matrix with rows
\begin{equation}\label{eq:R}
    R_{k} = i^{n_k}\mathcal{M}(\mathcal{A})_k, \quad 1 \leq k \leq M.
\end{equation}
It takes $16M$ bytes of memory to store the matrix $D$ (as each complex entry is 16 bytes). Furthermore, since $P_k$ only has $d$ nonzero entries, it takes approximately $3Md$ memory to store $R$ as a sparse matrix (1 byte each for data value, row position, and column position), which we accomplish using the scientific computation package \texttt{scipy}~\cite{2020SciPy-NMeth}. With this approach, storing $\mathcal{M}(\widetilde{\mathcal{A}})$ only requires around 3 MB.

There exist other approaches to storing $\widetilde{\mathcal{A}}$ that are more memory efficient. For instance, if we store each Pauli observable as a length-$n$ string, then the memory can be reduced to $\mathcal{O}(Mn)$. However, in this case, additional steps are needed to efficiently compute $\widetilde{\mathcal{A}}(\rho^t)$ and $\widetilde{\mathcal{A}}^\dagger(r^t)$.

\subsection{Comparison to the current art} \label{subsec:benchmarking}
\begin{figure*}[ht!]
    \centering
    \includegraphics[width=\linewidth]{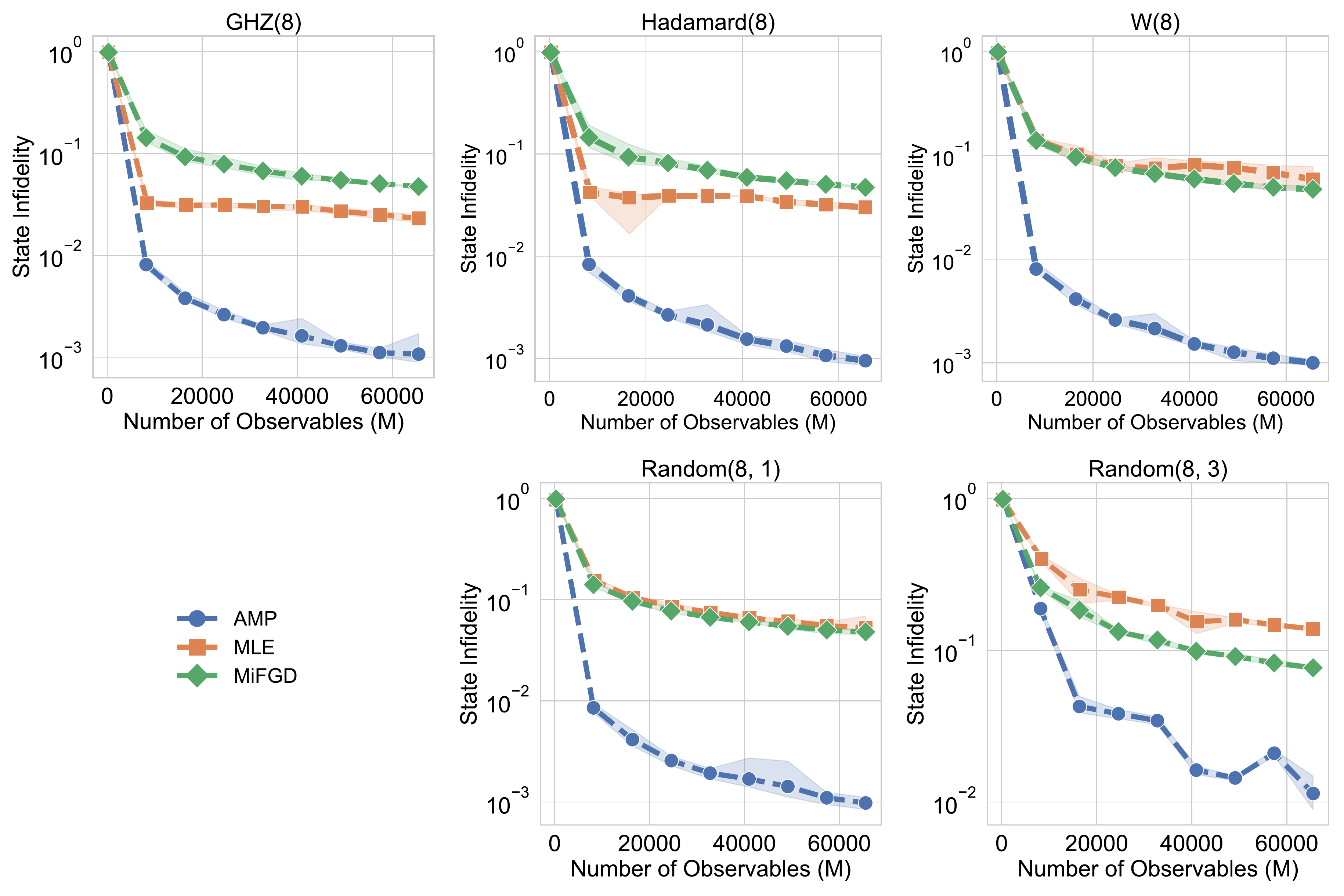}
    \caption{Comparison in reconstruction quality for 8-qubit states between approximate message passing (AMP), maximum likelihood estimation (MLE)~\cite{teo_incomplete_2012,teo_quantum-state_2011,teo_numerical_2013,sacchi_maximum-likelihood_2001,rehacek_diluted_2007,lvovsky_iterative_2004,jezek_quantum_2003,hradil_quantum-state_1997,goncalves_local_2012,glancy_gradient-based_2012,fiurasek_maximum-likelihood_2001,blume-kohout_hedged_2010,baumgratz_scalable_2013-1,aditi_rigorous_2025}, and momentum-inspired factored gradient descent (MiFGD)~\cite{kim_fast_2023}. We consider the GHZ, Hadamard, and W states, along with a random rank-1 and random rank-3 state. The shot count for each observable is fixed at $N=1024$. The $M$ observables are randomly sampled from the $d^2$ Pauli observables. Shaded regions indicate minimum and maximum state infidelity over 10 trials. AMP consistently outperforms both MLE and MiFGD, reducing the state infidelity by over an order of magnitude in almost all cases.}
    \label{fig:observable-comparison}
\end{figure*}

\begin{figure*}[ht!]
    \centering
    \includegraphics[width=\linewidth]{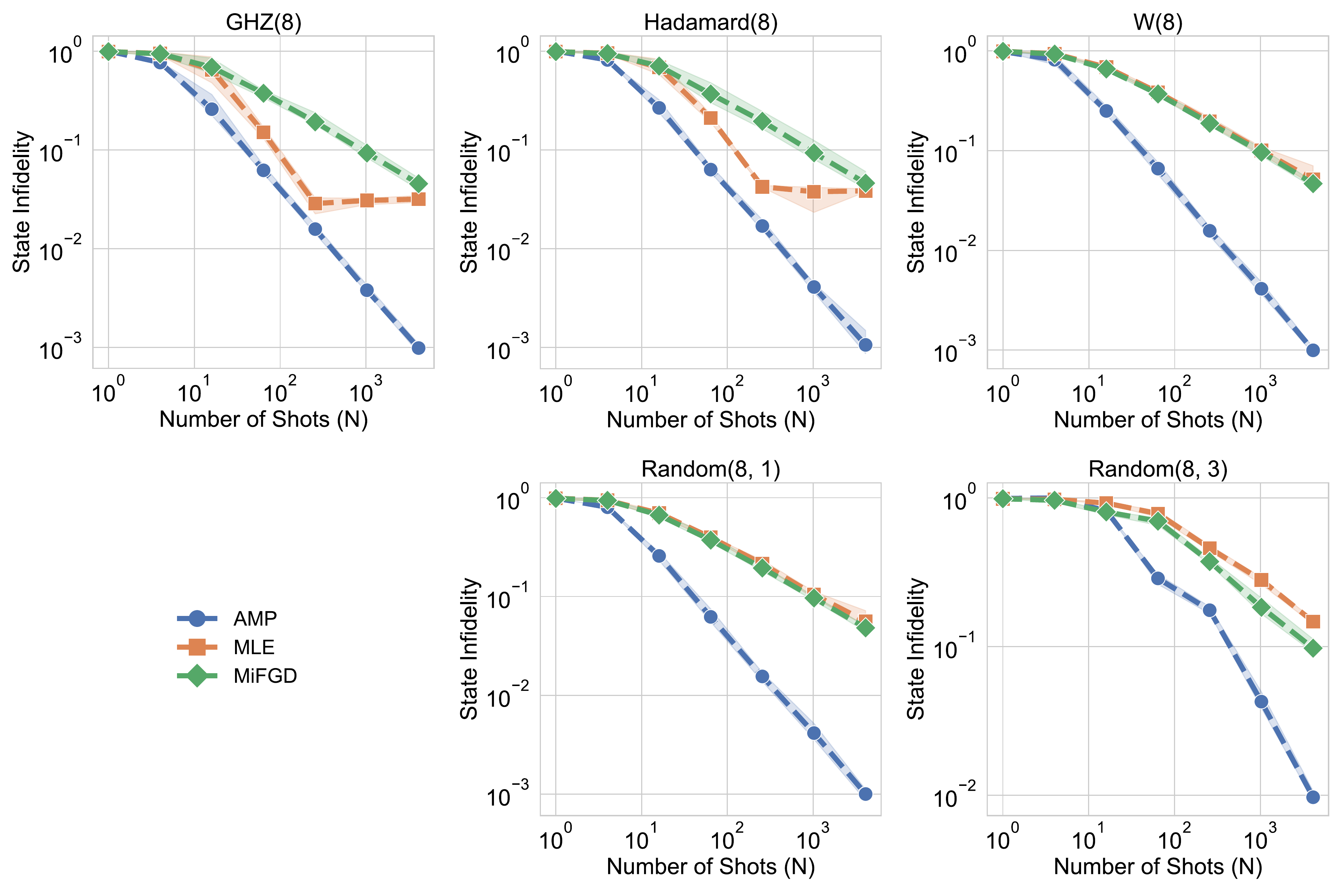}
    \caption{Comparison in reconstruction quality for 8-qubit states between approximate message passing (AMP), maximum likelihood estimation (MLE)~\cite{teo_incomplete_2012,teo_quantum-state_2011,teo_numerical_2013,sacchi_maximum-likelihood_2001,rehacek_diluted_2007,lvovsky_iterative_2004,jezek_quantum_2003,hradil_quantum-state_1997,goncalves_local_2012,glancy_gradient-based_2012,fiurasek_maximum-likelihood_2001,blume-kohout_hedged_2010,baumgratz_scalable_2013-1,aditi_rigorous_2025}, and momentum-inspired factored gradient descent (MiFGD)~\cite{kim_fast_2023}. We consider the GHZ, Hadamard, and W states, along with a random rank-1 and random rank-3 state. We fix $M = 16384$. The $M$ observables are randomly sampled from the $d^2$ Pauli observables. Shaded regions indicate minimum and maximum state infidelity over 10 trials. AMP consistently outperforms both MLE and MiFGD, reducing the state infidelity by over an order of magnitude in almost all cases.}
    \label{fig:shot-comparison}
\end{figure*}

We use MLE~\cite{teo_incomplete_2012,teo_quantum-state_2011,teo_numerical_2013,sacchi_maximum-likelihood_2001,rehacek_diluted_2007,lvovsky_iterative_2004,jezek_quantum_2003,hradil_quantum-state_1997,goncalves_local_2012,glancy_gradient-based_2012,fiurasek_maximum-likelihood_2001,blume-kohout_hedged_2010,baumgratz_scalable_2013-1,aditi_rigorous_2025} and MiFGD~\cite{kim_fast_2023} as baselines for our AMP-based approach to QST. The MLE minimizes the negative log-likelihood $\mathcal{L}$, given by
\begin{gather} \label{eq:mle-objective}
    \mathcal{L} = -\sum_{k=1}^M f_k \log p_k + (1-f_k) \log (1 -p_k),
\end{gather}
subject to the constraint that $\widehat{\rho} \in \mathcal{S}(\mathcal{H})$, where $f_k = (1+y_k)/2$ is the empirical frequency of the $+1$ eigenvalue of $P_k$ (see Eq.~\eqref{eq:fk}).
The minimization of $\mathcal{L}$ is a convex optimization problem that can be solved using Clarabel~\cite{goulart_clarabel_2024}, an interior-point solver wrapped by the convex modeling package \texttt{cvxpy}~\cite{Diamond-2016CVXPY}.

The MiFGD algorithm uses a factorization $\widehat{\rho} = UU^\dagger$, where $U \in \C^{d \times r}$ and $\rank\widehat{\rho} \le r$. The QST problem is then solved by the iteration
\begin{align}
    U^{t+1} &= Z^t - \eta \mathcal{A}^\dagger \left( \mathcal{A} \left(Z^t {Z^t}^\dagger \right) - y \right) Z^t, \label{eq:mifgd-iter-1} \\
    Z^{t+1} &= U^{t+1} + \mu \left(U^{t+1} - U^t \right), \label{eq:mifgd-iter-2}
\end{align}
for $0 \leq t < t_{\text{max}}$, where $U_0$ is randomly initialized and $Z_0 = U_0$. We set $\eta = 0.001$ and $\mu=3/4$ in accordance with ~\cite{kim_fast_2023}. The iteration given by~\eqref{eq:mifgd-iter-1} and~\eqref{eq:mifgd-iter-2} proceeds until either $t = t_\text{max} = 1000$ or $\norm{\rho^t - \rho^{t-1}}_F / \norm{\rho^t}_F < 10^{-4}$. We use a modification of the MiFGD implementation at~\cite{kim_mifgd_2021} that allows for $r > 1$. Since the rank of the quantum state $\rho^*$ to be reconstructed is \textit{a priori} unknown and neither the AMP nor the MLE algorithm have access to information about the rank of $\rho^*$, we set $r=5$ in the MiFGD reconstruction.

In Figure~\ref{fig:observable-comparison}, we compare the performance of AMP (blue circles), MLE (orange squares), and MiFGD (green diamonds) on the $8$-qubit states $\rho_{\text{GHZ}(8)}$, $\rho_{\text{Hadamard}(8)}$, $\rho_{\text{W}(8)}$, $\rho_{\text{Random}(8, 1)}$, and $\rho_{\text{Random}(8, 3)}$. We vary $M$ from 256 to 65536, which corresponds to full QST, and fix $N = 1024$. The state infidelity $1 - F(\rho^*, \widehat{\rho})$ is chosen to be the quality metric, as is common for QST. In the case that the reconstruction algorithm produces an error, e.g., due to numerical divergence, we report a state infidelity of 1.0.

Across the five different states and the range of observables, AMP consistently outperforms both MLE and MiFGD. For larger $M$, on all states except $\rho_{\text{Random}(8,3)}$, AMP improves the state infidelity by over an order of magnitude. The significance of this infidelity reduction  is exemplified in Table~\ref{tab:ghz-fid-comp}, which shows the four non-zero elements of $\rho_{\text{GHZ(8)}}$ for one state reconstruction where the state infidelity is $10^{-1}$ ($\widehat{\rho}_{0.1}$) and another where the state infidelity is $10^{-3}$ ($\widehat{\rho}_{0.001}$). With perfect reconstruction, all values should be $0.5$. We see large deviations, up to $10^{-1}$, between the entries of $\rho_{\text{GHZ}(8)}$ and $\widehat{\rho}_{0.1}$. For $\widehat{\rho}_{0.001}$, these deviations are all less than $10^{-3}$.

\begin{table}[h]
    \centering
    \begin{tabular}{|c|c|c|c|}
    \hline
         & $\rho_{\text{GHZ}(8)}$ & $\widehat{\rho}_{0.1}$ & $\widehat{\rho}_{0.001}$\\
         \hline
         $\rho_{1,1}$ & $0.5$ & $0.4010 + 0.0000i$ & $0.4992 + 0.0000i$ \\
         $\rho_{1,256}$ & $0.5$ & $0.4493 + 0.0006i$ & $0.4995 - 0.0006i$ \\
         $\rho_{256,1}$ & $0.5$ & $0.4493 - 0.0006i$ & $0.4995 + 0.0006i$ \\
         $\rho_{256,256}$ & $0.5$ & $0.5034 + 0.0000i$ & $0.4998 + 0.0000i$ \\
         \hline
    \end{tabular}
    \caption{Reconstructed entries of the $8$-qubit GHZ state $\rho_{\text{GHZ}(8)}$, with state infidelity $10^{-1}$ ($\widehat{\rho}_{0.1}$, exact infidelity $0.0986$) and $10^{-3}$ ($\widehat{\rho}_{0.001}$, exact infidelity $0.00097$). The real parts of the estimated entries for $\widehat{\rho}_{0.1}$ differ from the true values by no more than $10^{-1}$, and the real parts of the estimated entries for $\widehat{\rho}_{0.001}$ differ from the true values by no more than $10^{-3}$.}
    \label{tab:ghz-fid-comp}
\end{table}
In our numerical experiments, the runtime for a given algorithm (AMP, MLE, or MiFGD) for a fixed value of $M$ varied by up to an order of magnitude. We attribute this variation in runtime to interference from other compute jobs running on the same cluster node. The runtimes for the three algorithms (AMP, MLE, and MiFGD) are also within an order of magnitude of each other, which suggests that the runtime for all three low-rank QST algorithms is comparable.

We are also interested in studying the effect of shot noise on reconstruction quality for each QST algorithm. In Figure~\ref{fig:shot-comparison}, we fix $M = 16384$ ($= 0.25d^2$) and vary $N$. We see that AMP consistently matches or outperforms both MLE and MiFGD on all five states we consider, with larger reductions in state infidelity as $N$ is increased. The AMP-based QST algorithm appears to exhibit Heisenberg-limited ($\mathcal{O}(1/N)$) precision scaling in the state infidelity, whereas the MLE- and MiFGD-based QST algorithms appear to only achieve $\mathcal{O}(1/\sqrt{N})$ precision scaling. We leave the investigation of the origin of this scaling for future work.

\begin{figure}
    \centering
    \includegraphics[width=\linewidth]{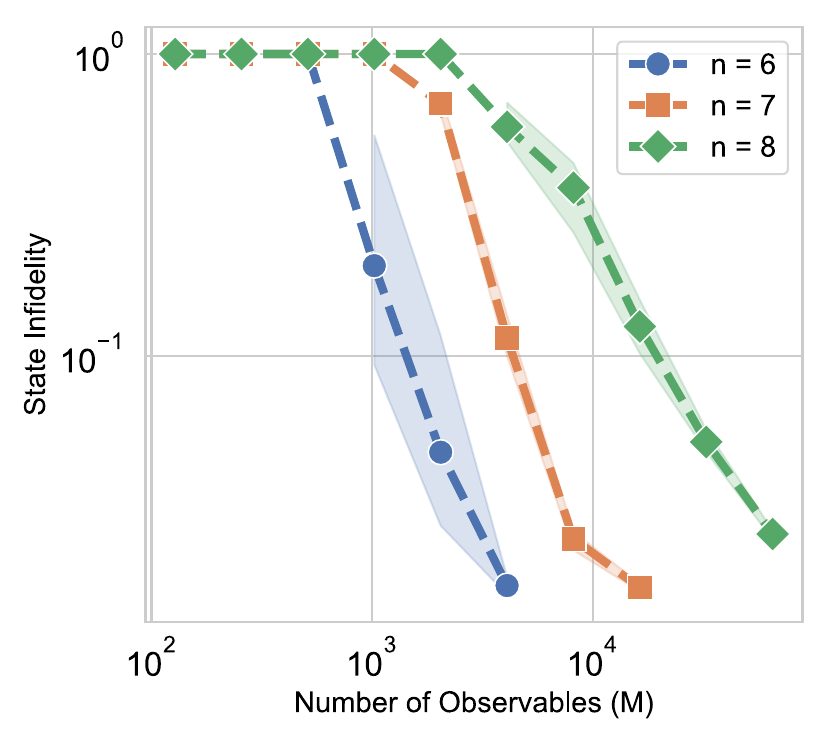}
    \caption{Recovering rank-$n$ states with AMP. We run AMP on random $n$-qubit states of rank $n$ for $n=6$, $n=7$, and $n=8$ with $N=4096$ shots per observable. As $n$ increases, so does the number of observables required to recover the random state. However, the fraction $M/4^n$ required to reach a reconstruction state fidelity of $10^{-1}$ decreases with increasing $n$. Shaded regions show the maximum and minimum state infidelity over 10 trials.}
    \label{fig:qubit-scaling}
\end{figure}

\subsection{Reconstructing higher-rank states} \label{subsec:higher-rank-states}

In the previous subsection, we mainly considered the reconstruction of pure states, i.e., $\rank \rho^* = 1$. However, in the presence of device noise, the preparation of a target state $\rho_\text{target}$ with $\rank \rho_\text{target} = 1$ can be corrupted by noise and instead produce a noisy state $\rho^*$ whose rank may be greater than unity. In the case of a coherent error channel $\mathcal{E}_{\text{coher}} \colon \H^{d \times d} \rightarrow \H^{d \times d}$ given by
\begin{equation} \label{eq:def-coherent-error}
    \mathcal{E}_{\text{coher}}[\rho] = C \rho C^\dagger,
\end{equation}
where $C$ is a unitary operator~\cite{nielsen_quantum_2012} on the Hilbert space $\mathcal{H}$, we have $\rank (\mathcal{E}_{\text{coher}}[\rho_\text{target}]) = 1$. On the other hand, for a depolarizing channel
$\mathcal{E}_\text{depol}\colon \H^{d \times d} \rightarrow \H^{d \times d}$ given by
\begin{equation} \label{eq:def-depol-error}
    \mathcal{E}_\text{depol}[\rho] = (1-\epsilon)\rho + \frac{\epsilon}{d}\mathbb{I}_d,
\end{equation}
where $0 < \epsilon \leq 1$, the resulting state is full-rank: $\rank (\mathcal{E}_{\text{depol}}[\rho_\text{target}]) = d$. We are interested in a noise model that lies somewhere between these two extremes. 

To find such a noise model, we consider a photonic quantum device where the only error channels are bit-flips, phase-flips, and photon loss. As we prove in Appendix~\ref{appen:photon-error-model}, this implies that $\rank \rho^* \le 6n+1$. Motivated by this photonic noise model, we consider states of the form $\rho^* = \rho_{\text{Random}(n,n)}$. Figure~\ref{fig:qubit-scaling} shows the state infidelity $1 - F(\rho^*, \widehat{\rho})$ for $n = 6$, $7$, and $8$. We vary $M$ and fix $N = 4096$. The state infidelity for $\rho_{\text{Random}(8,8)}$ is higher compared to $\rho_{\text{Random}(8,1)}$ and $\rho_{\text{Random}(8,3)}$ (Figure~\ref{fig:observable-comparison}), despite the increased shot count. Full tomography ($M = 4096$) on the $\rho_{\text{Random}(6,6)}$ state reduces the state infidelity by half an order of magnitude compared to $M = 2048$. This behavior is notably different from that observed in the results shown in Figure~\ref{fig:observable-comparison}. There, although the state infidelity decreases with $M$, the outperformance is smaller. Even with full tomography, we do not reach an infidelity of $10^{-2}$ for any of the rank-$n$ random states.

The behavior shown in Figure~\ref{fig:qubit-scaling} is consistent with the $\mathcal{O}(rdn^2) = \mathcal{O}(dn^3)$ scaling in the number of observables necessary to perform QST, since for small $n$, there is minimal separation between $dn^3$ (low-rank QST) and $d^2$ (full QST). The fact that full tomography---or at least $M = 0.5d^2$ tomography---is needed to obtain state infidelity below $0.1$ suggests that for $n=6$, $7$, and $8$, it may not be appropriate to consider these states as low-rank. For larger $n$, the separation between $dn^3$ and $d^2$ is larger, and therefore the reconstruction of $\rho_{\text{Random}(n,n)}$ may benefit more from a low-rank QST algorithm.

\section{Experimental QST with AMP} \label{sec:experimental-results}
The end goal of any QST technique is to apply it to a real quantum device and characterize the state preparation of a target state $\rho_{\text{target}}$ specified by the circuit $C_{\rho_{\text{target}}}$. Due to device errors,  $C_{\rho_{\text{target}}}$ prepares a noisy state $\rho^*$ instead of $\rho_{\text{target}}$. The task of QST is to reconstruct the prepared state $\rho^*$. Running QST experiments on hardware creates additional considerations compared to the numerical simulations reported in the previous section, which arise from quantum resource requirements and the impact of device errors on QST reconstruction. In the following subsections, we elaborate on these considerations and describe the modifications we make to our QST workflow in order to address them.

\subsection{Quantum runtime} \label{subsec:settings-comp}

\begin{table*}[ht!]
    \centering
    \begin{tabular}{|c||c|c|c||c|c|c||c|c|c||c|c|c|}
        \hline
        & \multicolumn{12}{c|}{$M/d^2$} \\
        \cline{2-13}
         & \multicolumn{3}{c||}{0.25} & \multicolumn{3}{c||}{0.5} & \multicolumn{3}{c||}{0.75} & \multicolumn{3}{c|}{1.0}\\
        \hline
        $n$ & $M$ & $T$ & $T / M$ & $M$ & $T$ & $T / M$ & $M$ & $T$ & $T / M$ & $M$ & $T$ & $T / M$ \\
        \hline
        3 & 16 & 3 & 18.8\% & 32 & 7 & 21.1\% & 48 & 14 & 29.1\% & 64 & 27 & 42.2\%\\
        4 & 64 & 6 & 9.3\% & 128 & 16 & 12.6\% & 192 & 36 & 18.7\% & 256 & 81 & 31.6\% \\
        5 & 256 & 13 & 4.9\% & 512 & 37 & 7.3\% & 768 & 90 & 11.8\% & 1024 & 243 & 23.7\% \\
        6 & 1024 & 27 & 2.6\% & 2048 & 86 & 4.2\% & 3072 & 227 & 7.4\% & 4096 & 729 & 17.8\% \\
        7 & 4096 & 59 & 1.4\% & 8192 & 200 & 2.4\% & 12288 & 559 & 4.6\% & 16384 & 2187 & 13.3\% \\
        8 & 16384 & 128 & 0.8\% & 32768 & 458 & 1.4\% & 49152 & 1369 & 2.8\% & 65536 & 6561 & 10.0\% \\
        9 & 65536 & 281 & 0.4\% & 131072 & 1051 & 0.8\% & 196608 & 3326 & 1.7\% & 262144 & 19683 & 7.5\% \\
        10 & 262144 & 614 & 0.2\% & 524288 & 2406 & 0.5\% & 786432 & 8038 & 1.0\% & 1048576 & 59049 & 5.6\% \\
        \hline
    \end{tabular}
    \caption{Sampling circuits based on measurement settings instead of observables. Number of measurement settings $T$ required to obtain $M = 0.25d^2$, $M = 0.5d^2$, $M = 0.75d^2$, $M = d^2$ observables for $n=3$ to $n=10$ (note that $T=3^n$ when $M=d^2=4^n$). Measurement settings are randomly sampled from $\{X,Y,Z\}^n$ without replacement. We ran 100 trials, but we do not include the standard deviation as it is always small compared to $T$.}
    \label{tab:obs-for-meas}
\end{table*}

\begin{figure}
    \centering
    \includegraphics[width=\linewidth]{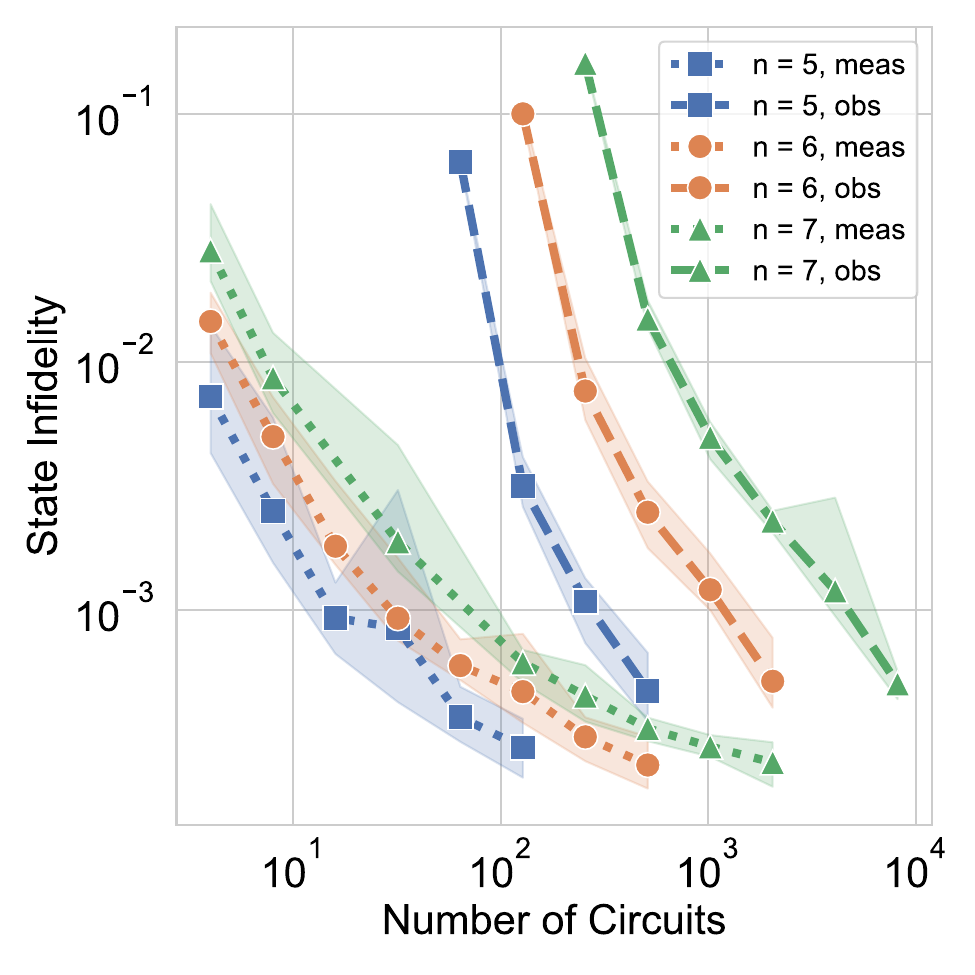}
    \caption{Comparison between sampling circuits based on measurement settings and based on observables. We use AMP to recover a random rank-1 state on $n$ qubits for $n=4,5,6$ with $N=4096$ shots per circuit. When sampling circuits based on measurement settings instead of observables, each circuit can be used to estimate the expectation values of $d$ observables, which reduces the number of circuits that need to be run in order to perform QST. We randomly sample circuits based on either measurement settings or observables and report the average state infidelity. Shaded regions show the minimum and maximum state infidelity. We observe a large reduction in the number of circuits needed to perform QST to a desired fidelity when circuits are sampled based on measurement settings instead of observables.}
    \label{fig:sampling}
\end{figure}

In simulation, the only resource concern for QST is the time and memory required to run the QST algorithm on a classical computer. As the system size $d$ and the number of observables $M$ increase, both the runtime and memory tend to increase. Hence, efficiently simulating QST is only a matter of reducing the classical resource requirements. However, for experimental QST, we must also consider the cost of collecting from a quantum processing unit (QPU) the data 
used to generate the measurement vector $y$.

In order to estimate the expectation value of each Pauli observable $P_k$, we must run a quantum circuit, say $C_k$, whose empirical outcome distribution, i.e., shot data, can be used to estimate $\Tr [P_k \rho^*]$. The simplest such approach is to define $C_k$ as the composition of the state preparation circuit $C_{\rho_\text{target}}$ with a measurement in the $P_k$ basis. Denoting the $j$-th letter in the Pauli string $P_k$
by $(P_k)_j$, measuring in the $P_k$ basis corresponds to measuring in the $(P_k)_j$ basis on the $j$-th qubit for all $1 \leq j \leq n$. In this approach, if $(P_k)_j = I$, then qubit $j$ is not measured. Thus, for $M$ observables, $M$ unique circuits are required.

Reducing the number of circuits required for QST is important for two reasons. First, the QPU runtime required for QST scales linearly in the number of circuits used. QPU runtime is a constrained and expensive resource, and it is therefore important to be efficient in the utilization of this runtime. Second, from a scientific perspective, increased runtime means that there may be stronger effects from device drift~\cite{proctor_detecting_2020}. When performing QST, we assume that each time we run $C_{\rho_\text{target}}$, it prepares the same noisy state $\rho^*$. This assumption is not realized in practice due to imperfect control in the quantum computer, but it is stretched further when the timescale over which the QST data are collected is on the order of a calibration cycle, which can happen for moderately sized experiments---see Section~\ref{subsec:ibmq}. Over this calibration cycle timescale, it is possible that the components involved in the state preparation drift substantially, in which case the observable data we estimate using all such circuits $C_k$ are not well identified with a unique $\rho^*$. While our AMP algorithm may indeed produce an estimate $\widehat{\rho}$, it is unclear what state is being estimated by $\widehat{\rho}$. Reducing the number of circuits reduces the QPU runtime, which means that the estimate $\widehat{\rho}$ corresponds to a more well-defined state $\rho^*$.

In order to measure $M$ Pauli observables using fewer than $M$ circuits, we define a \emph{measurement setting} $S \in \{X,Y,Z\}^n$ as a specification for which basis to measure in on each qubit. If we execute a circuit with this measurement setting, then the expectation value of any observable that differs from $S$ only by replacing some of the letters in $S$ by $I$ can be obtained by marginalizing over the qubits where $I$ is measured. See Appendix~\ref{appen:meas-to-obs} for more details. Consequently, the measurement setting $S$ enables the estimation of the expectation values of $2^n$ Pauli observables, instead of one Pauli observable as in the simple approach. For instance, the circuit corresponding to the measurement setting $XY$ allows us to estimate the expectation values of $X \otimes Y$, $X \otimes I$, $I \otimes Y$, and $I \otimes I$.

However, the number of observables is not uniquely determined by the number of measurement settings. Let $P_{S_1}$ be the set of Pauli observables whose expectation values we can estimate using measurement setting $S_1$, and $P_{S_2}$ be the set of Pauli observables whose expectation values we can estimate using measurement setting $S_2$. Clearly $P_{S_1}$ and $P_{S_2}$ are not disjoint---both contain $I^{\otimes n}$, and they may contain other observables in their intersection as well. Consequently, when we randomly sample $T$ measurement settings from $\{X,Y,Z\}^n$, where $1 < T < 3^n - 1$, the number of observables $M$ that we can estimate will vary. 

In Table~\ref{tab:obs-for-meas}, we report the average number of randomly sampled measurement settings $T$ needed to estimate $M$ observables for $M/d^2 = [0.25, 0.5, 0.75, 1]$ and $n = 3$ to $n=10$ across 100 trials. While there is variation among trials, the standard deviation in the number of measurement settings required is always much smaller than $T$. We also report $T/M$ as a percentage. For fixed $M/d^2$, as $n$ increases, $T/M$ decreases,  i.e., $T$ grows slower than $M$. Hence, this measurement setting sampling strategy becomes more efficient for larger $n$. For fixed $n$, as $M/d^2$ increases, the efficiency of sampling measurement settings decreases. However, since we apply QST to low-rank states, we have $M < d^2$, and thus we benefit from increased efficiency compared to full QST where $M/d^2 = 1$. Even when $M = d^2$, running $3^{10} = 59049$ circuits instead of $4^{10} = 1048576$ circuits is almost a twenty-fold reduction in QPU runtime.

In Figure~\ref{fig:sampling}, we show how the approaches of generating circuits based on observables and based on measurement settings compare when reconstructing a state $\rho^* = \rho_{\text{Random}(n,1)}$ for $n = 5$, $6$, $7$. We fix $N=4096$ and report the state infidelity $1 - F(\rho^*, \widehat{\rho})$. When generating circuits based on measurement settings, we are able to reconstruct $\rho^*$ with lower infidelity using fewer circuits than when we sample circuits based on observables. This advantage is more pronounced as $n$ increases.

As discussed in Section~\ref{subsec:ibmq}, we employ the measurement-setting-based circuit creation approach when running QST experiments on IBM Kingston, in order to reduce the QPU runtime needed to perform tomography.

\subsection{Predicting state preparation fidelity in the presence of noise} \label{subsec:pred-fid-under-noise}
\begin{figure}
    \centering
    \includegraphics[width=\linewidth]{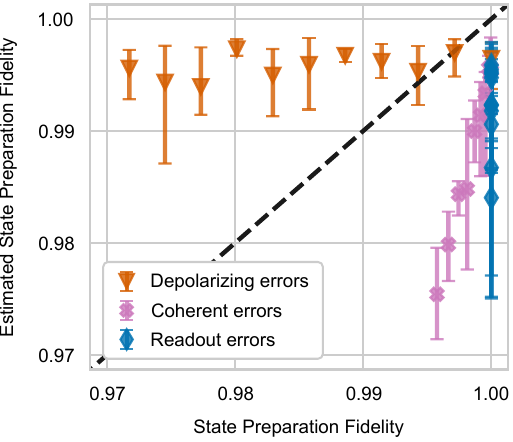}
    \caption{Predicting the fidelity of state preparation using AMP-based QST. We reconstruct a $5$-qubit GHZ state using $T=37$ measurement settings and $N=1024$ shots per circuit. Reported is the estimated state fidelity $F(\rho_{\text{target}}, \widehat{\rho})$ between the target and reconstructed state versus the true state fidelity $F(\rho_{\text{target}}, \rho^*)$ between the target and prepared state in the presence of depolarizing, coherent, and readout errors, simulated using Qiskit Aer. In the presence of depolarizing errors, $F(\rho_{\text{target}}, \widehat{\rho})$ overestimates $F(\rho_{\text{target}}, \rho^*)$. In the presence of coherent errors or readout errors, $F(\rho_{\text{target}}, \widehat{\rho})$ underestimates $F(\rho_{\text{target}}, \rho^*)$. Error bars are the maximum and minimum estimated state fidelity.}
    \label{fig:pred-fid-under-noise}
\end{figure}

In our numerical results, the reported state fidelity $F(\rho^*, \widehat{\rho})$ (or infidelity, $1 - F(\rho^*, \widehat{\rho})$) served to quantify the quality of the reconstruction. In the case of experimentally deployed QST, we are instead interested in estimating the state preparation fidelity given by $F(\rho_{\text{target}}, \rho^*)$, where $\rho_{\text{target}}$ is the state we designate to be prepared and $\rho^*$ is the noisy outcome of this state preparation on a quantum device. However, we do not have access to $\rho^*$ outside of numerical simulations, and therefore we instead compute $F(\rho_{\text{target}}, \widehat{\rho})$ as a proxy for $F(\rho_{\text{target}}, \rho^*)$. Under the assumption that $\widehat{\rho} \approx \rho^*$, we have $F(\rho_{\text{target}}, \widehat{\rho}) \approx F(\rho_{\text{target}}, \rho^*)$. In this section, we use noise simulations to investigate the validity of this approximation in the presence of physically realistic noise channels. We consider depolarizing errors~\eqref{eq:def-depol-error}, coherent errors~\eqref{eq:def-coherent-error}, and readout errors. Readout error is described by a classical error channel, in which a bit $b$ flips to $1-b$ with probability $q$.

To test the effect of each kind of error---depolarizing, coherent, and readout---we numerically simulate the reconstruction of the $5$-qubit GHZ state $\rho_{\text{GHZ}(5)}$ in the presence of noise. We run these noise simulations in \texttt{qiskit\_aer}~\cite{javadi-abhari_quantum_2024} to obtain $\rho^*$ and collect shot data. For these simulations, we transpile the measurement circuits $C_k$ to a $[X, SX, RZ, CZ]$ gate set using the \texttt{qiskit}~\cite{javadi-abhari_quantum_2024} transpiler. These gates are given by~\cite{nielsen_quantum_2012}:
\begin{align}
    X &= i RX(\pi), \\
    SX &= e^{i\pi/4} RX(\pi/2), \\
    RX(\theta) &= \begin{pmatrix}
        \cos(\theta/2) & -i \sin(\theta/2) \\
        -i \sin(\theta/2) & \cos(\theta/2)
    \end{pmatrix},\\
    RZ(\theta) &= \begin{pmatrix}
        e^{-i \theta /2} & 0 \\
        0 & e^{i \theta /2}
    \end{pmatrix},\\
    CZ &= \begin{pmatrix}
        1 & 0 & 0 & 0 \\
        0 & 1 & 0 & 0 \\
        0 & 0 & 1 & 0 \\
        0 & 0 & 0 & -1
    \end{pmatrix}, \label{eq:def-cz-gate}
\end{align}
where $\theta \in \R$. We sample $T=37$ measurement settings, which yield approximately 512 ($0.5d^2$) Pauli observables (see Table~\ref{tab:obs-for-meas}), and take $N=1024$ shots for each circuit. Figure~\ref{fig:pred-fid-under-noise} shows the results of these simulations, comparing $F(\rho_{\text{target}}, \widehat{\rho})$---which estimates the state preparation fidelity---to the true state preparation fidelity $F(\rho_{\text{target}}, \rho^*)$. For the depolarizing noise model (orange triangles), we apply depolarizing errors to all $X$, $SX$, and $CZ$ gates, and let $\epsilon$ range from $0$ to $0.0005$. The coherent error model (purple crosses) applies a coherent overrotation by $\theta$ to each $SX$ gate (so $SX$ is replaced by $SX \cdot RX(\theta)$), where $\theta$ ranges from $0$ to $0.09$. For the readout error model (blue diamonds), we vary $q$ from $0$ to $0.05$.

In the presence of depolarizing noise, the state fidelity $F(\rho_{\text{target}}, \widehat{\rho})$ overpredicts the state preparation fidelity $F(\rho_{\text{target}}, \rho^*)$. This behavior can be understood by noting that depolarizing errors uniformly lift the zero singular values of the density matrix $\rho_\text{target}$, and therefore the PSVT denoiser~\eqref{eq:def-psvt}---which is based on singular value thresholding---may treat the depolarizing noise as statistical noise. Hence, the AMP algorithm may reconstruct $\rho_{\text{target}}$ instead of $\rho^*$, which causes $F(\rho_{\text{target}}, \widehat{\rho})$ to overestimate $F(\rho_{\text{target}}, \rho^*)$.

For the coherent noise model, $F(\rho_{\text{target}}, \widehat{\rho})$ underpredicts $F(\rho_{\text{target}}, \rho^*)$. Coherent errors do not change the rank of $\rho_{\text{target}}$, and therefore the previously described conflation of device noise with statistical noise should not present itself. However, a new problem arises. In order to measure in either the $X$ or $Y$ basis, we must first perform a rotation and then a computational basis measurement. These rotations involve $SX$ gates, and hence they are also affected by the coherent errors. Thus, we reconstruct a quantum state other than $\rho^*$ that may not be correlated with $\rho_{\text{target}}$, and therefore$F(\rho_{\text{target}}, \widehat{\rho})$ may underestimate $F(\rho_{\text{target}}, \rho^*)$. This behavior is observed in Figure~\ref{fig:pred-fid-under-noise}.

For the readout error model, $F(\rho_{\text{target}}, \widehat{\rho})$ underpredicts $F(\rho_{\text{target}}, \rho^*)$. Indeed, since readout error is a classical error channel, the state preparation is noiseless: $\rho^* = \rho_{\text{target}}$. Hence, $F(\rho_{\text{target}}, \rho^*) = 1$, while $F(\rho_{\text{target}}, \widehat{\rho}) < 1$ due to the corruption of the shot data by readout errors.

From these noise simulations, it is clear that one must exercise caution when interpreting the results of a QST experiment, as the reconstructed state $\widehat{\rho}$ does not always produce an accurate fidelity prediction. Moreover, whether the fidelity of state preparation is underpredicted or overpredicted depends on the details of the noise model. We note that the effect of coherent errors on measurements could potentially be reduced using composite pulse sequences~\cite{brown_arbitrarily_2004}. Likewise, readout error could potentially be mitigated using expectation minimization~\cite{chandramouli_statistical_2025} or another readout error mitigation technique. We leave the analysis of such techniques and their interplay with AMP-based QST for future work.

\subsection{Tomography experiments on IBM Q} \label{subsec:ibmq}

\begin{figure}
    \centering
    \includegraphics[width=\linewidth]{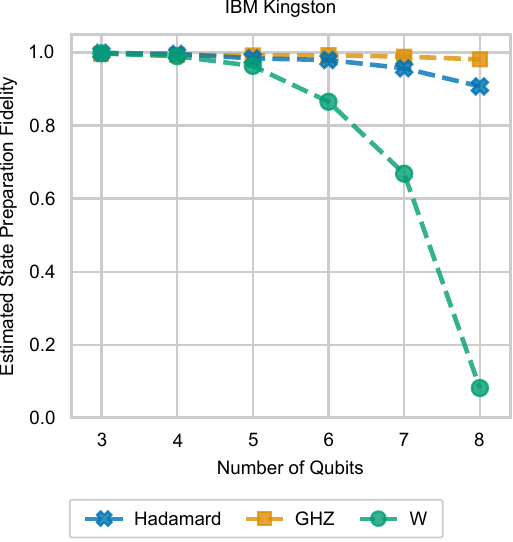}
    \caption{Running AMP-based QST  on IBM Kingston. We prepare the GHZ, Hadamard, and W state for $n=3$ to $n=8$ qubits and reconstruct the noisy states using AMP. The number of measurement circuits was chosen to estimate the expectation values of $0.75 \cdot 4^n$ observables ($n=3$) and $0.5 \cdot 4^n$ observables ($n=4$ to $n=8$). We then compute the state fidelity $F(\rho^*, \widehat{\rho}$) between the target pure state and the reconstructed density matrix. Despite requiring more two-qubit gates and a deeper circuit, the predicted GHZ state preparation fidelity is higher than that of the Hadamard state, which suggests that the AMP reconstruction is affected by a form of device noise.}
    \label{fig:ibmq-experiments}
\end{figure}

We ran tomography experiments for $\rho_{\text{GHZ}(n)}$, $\rho_{\text{Hadamard}(n)}$, and $\rho_{\text{W}(n)}$ for $n=3$ to $n=8$ qubits on IBM Kingston. The number of measurement circuits was chosen to achieve $M = 0.5d^2$ (see Table~\ref{tab:obs-for-meas}) for $n = 4$ to $n = 8$; for $n = 3$, we ran 14 circuits to achieve $M = 0.75d^2$. This consideration is the same as in Section~\ref{subsec:higher-rank-states}: $n=3$ ($d=8$) is small enough that close to full tomography is required for reconstruction. These experiments used a total of 12 minutes of QPU runtime and 811 circuits; without the measurement setting optimization described in Section~\ref{subsec:settings-comp}, 43696 circuits and an estimated 11 hours of QPU runtime would have been required.

At a high level, there are two factors that have a prominent effect on the state preparation fidelity. The first is the depth of the state preparation circuit: longer circuits tend to accumulate more error since there are more operations where errors can occur. The second is the number of two-qubit gates in the state preparation circuit, as two-qubit gates tend to be noisier than single-qubit gates~\cite{gidney_fault-tolerant_2021,orourke_compare_2024,otxoa_spinhex_2025}. The native two-qubit gate for IBM Kingston is the $CZ$ gate~\eqref{eq:def-cz-gate}. In Table~\ref{tab:transpiled-circ-depths}, we report the transpiled depth and $CZ$ gate count for the $\rho_{\text{GHZ}(n)}$, $\rho_{\text{Hadamard}(n)}$, and $\rho_{\text{W}(n)}$ state preparation circuits. The Hadamard circuits only contain a Hadamard gate on each qubit and have no entangling gates; thus, they are constant-depth circuits. For the GHZ state preparation circuits, both the circuit depth and $CZ$ count grow linearly with $n$. Both the circuit depth and $CZ$ count for the W circuits grow approximately exponentially in $n$, doubling each time $n$ increases by 1. 

\begin{table}[h]
    \centering
    \begin{tabular}{|c||c|c||c|c||c|c|}
    \hline
         & \multicolumn{2}{c||}{$\rho_{\text{Hadamard}(n)}$} & \multicolumn{2}{c||}{$\rho_{\text{GHZ}(n)}$} & \multicolumn{2}{c|}{$\rho_{\text{W}(n)}$} \\
         \hline
         $n$ & Depth & $CZ$ count & Depth & $CZ$ count & Depth & $CZ$ count \\
         \hline
         3 & 3 & 0 & 9 & 2 & 30 & 7 \\
         4 & 3 & 0 & 12 & 3 & 63 & 13\\
         5 & 3 & 0 & 15 & 4 & 147 & 34 \\
         6 & 3 & 0 & 18 & 5 & 314 & 106 \\
         7 & 3 & 0 & 21 & 6 & 692 & 193\\
         8 & 3 & 0 & 24 & 7 & 1378 & 454 \\
         \hline
    \end{tabular}
    \caption{Transpiled circuit depth and $CZ$ count for Hadamard, GHZ, and W state preparation circuits for $n = 3$ to $n=8$ on IBM Kingston. The Hadamard circuit is constant depth and has no entangling gates. The circuit depth and number of $CZ$ gates both grow linearly in $n$ for the GHZ state preparation circuits. The depth of the W state preparation circuit approximately doubles each time $n$ increases by 1, as does the $CZ$ count.}
    \label{tab:transpiled-circ-depths}
\end{table}

In Figure~\ref{fig:ibmq-experiments}, we show the predicted state preparation fidelity $F(\rho_\text{target}, \widehat{\rho})$ for each state based on our tomography experiments. The predicted state preparation fidelity remains above 0.9 for the Hadamard states (blue crosses) and above 0.98 for the GHZ states (orange squares). Given the depth and two-qubit gate count for the Hadamard and GHZ states (Table~\ref{tab:transpiled-circ-depths}), it seems unlikely that the true state preparation fidelity for the Hadamard circuits would be lower than that of the GHZ states. Based on our noise simulations in Section~\ref{subsec:pred-fid-under-noise}, we conjecture that the AMP reconstruction may be underpredicting the state preparation fidelity for Hadamard states (e.g., due to coherent errors) or overpredicting the state prearation fidelity for GHZ states (e.g., due to depolarizing errors). For the W state (green circles), $F(\rho_\text{target}, \widehat{\rho})$ falls off rapidly with $n$, which is consistent with the exponential increase in both circuit depth and $CZ$ count for the W state preparation circuits. 

\section{Discussion}\label{sec:disc}
We have demonstrated how approximate message passing (AMP) can be applied to the quantum state tomography (QST) problem for low-rank states and improve the recovery quality compared to other compressed sensing techniques. We also demonstrated an experimental application of AMP-QST, with consideration for the quantum processing unit (QPU) runtime cost and the effects of device noise on the reliability of the reconstruction.

There are multiple open questions that remain to be addressed. The first is if the AMP formalism can be expanded to provide rigorous performance guarantees for our QST algorithm. As described in Section~\ref{subsec:AMP_QST}, the QST sensing map does not meet the technical requirements under which state evolution has been proven for AMP, but it is possible that an extension of the formalism will address our QST use case. Such a theoretical advancement may explain the Heisenberg-limited scaling observed in Fig~\ref{fig:shot-comparison}, and may also admit better algorithm performance. It may also be possible to improve the performance of the AMP algorithm by changing the denoiser, either by modifying the PSVT denoiser \eqref{eq:def-psvt}---such as with Frobenius-norm-optimal projection---or by using a different denoising function. We have not attempted to optimize the convergence speed, and our numerical simulations suggest that setting $\lambda = 0.01$ and $t_{\text{max}} = 2000$ is conservative. Beyond the $DR$ factorization (see \eqref{eq:D}, \eqref{eq:R}), we have also not attempted to optimize the memory footprint of our AMP algorithm. Improving these resource requirements, runtime and memory, will increase the practical utility of our AMP approach to QST.

We are optimistic about the application of AMP to other tomography problems where the large system limit (Condition~1) also applies. There are other classes of interesting states, e.g., matrix product states, where AMP may yield a recovery advantage. Moreover, it may be possible to apply AMP to quantum process tomography~\cite{Poyatos-1997CompleteQPT,Surawy-2022Projected,Ahmed-2023LearningKraus}, quantum measurement tomography~\cite{Lundeen-2009Tomography,Chen-2019detector,Zambrano-2025Fast}, and gate set tomography~\cite{Nielsen-2021GST}, by leveraging the structure inherent in each problem. Compressed sensing techniques have previously been applied to these tomography tasks---see, e.g.,~\cite{Kliesch-2019Guaranteed,Quiroga-2023QPT,Volya-2024Fast}---and thus we anticipate that AMP may provide a performance improvement there as well. 

\section*{Data availability}
The data that supports the findings of this article cannot be made publicly available because they contain commercially sensitive information. The data are available upon reasonable request from the authors.

\section*{Acknowledgments}
We thank Cindy Rush and
Ramji Venkataramanan for helpful 
discussions about denoising low-rank matrices, and Dima Farfurnik for discussions on photon error models.
B.N.B. and D.B. were supported in part by the U.S. Department of Energy, Advanced Scientific Computing Research, under contract number DE-SC0025384. 
We acknowledge the computing resources provided by North Carolina State University High Performance Computing Services Core Facility (RRID:SCR\_022168). N.S. thanks Riley Murray, Corey Ostrove, and Rich Lehoucq for helpful conversations. N.S. and K.C. acknowledge helpful conversations with Jonathan Cranford. We acknowledge the use of IBM Quantum services for this work. The views expressed are those of the authors, and do not reflect the official policy or position of IBM or the IBM Quantum team.

\bibliography{literature,siekierski_amp,siekierski_qst}
\bibliographystyle{apsrev4-1}

\appendix

\section{A photonic error model} \label{appen:photon-error-model}
In this appendix, we show that if we consider a noise model with bit-flips, phase-flips, and photon loss, then the rank of the resulting state is bounded linearly in $n$. Bit-flips and phase-flips are special cases of coherent error channels as in \eqref{eq:def-coherent-error}, namely
\begin{equation} \label{eq:bit-phase-flip}
    \mathcal{E}_{\text{bit},i}[\rho] = X_i \rho X_i, \quad
    \mathcal{E}_{\text{phase},i}[\rho] = Z_i \rho Z_i,
\end{equation}
for $1\le i\le n$, where 
\begin{equation}
\begin{split}
X_i &= I^{\otimes (i-1)} \otimes X \otimes I^{\otimes (n-i)}, \\
Z_i &= I^{\otimes (i-1)} \otimes Z \otimes I^{\otimes (n-i)},
\end{split}
\end{equation}
denote the corresponding Pauli matrices acting on the $i$th qubit
(see \eqref{pauli-1}, \eqref{pauli-2}).

Let us now consider a photon loss error channel. A loss of the first qubit can be represented as a partial trace of $\rho$ over the first qubit combined with depolarization of the first qubit (cf.\ \eqref{eq:def-depol-error}). Explicitly, we can write a $2^n \times 2^n$ density matrix $\rho$ in the block form
\begin{align*}
\rho &= \begin{pmatrix}
    A & B \\
    C & D
\end{pmatrix} \\
&= \ket0\bra0\otimes A + \ket0\bra1\otimes B + \ket1\bra0\otimes C + \ket1\bra1\otimes D,
\end{align*}
where the blocks $A,B,C,D$ are of size $2^{n-1} \times 2^{n-1}$.
Then a loss of the first qubit corresponds to the channel
\begin{equation}\label{eq:loss1}
\begin{split}
    \mathcal{E}_{\text{loss},1}[\rho] 
    &= \frac12 \mathbb{I}_2 \otimes A + \frac12 \mathbb{I}_2 \otimes D \\
    &= \frac12 \begin{pmatrix}
    A+D & 0 \\
    0 & A+D
\end{pmatrix}.
\end{split}
\end{equation}
When $\rank\rho=1$, all rows of $A$ are scalar multiples of each other, and all rows of $D$ are scalar multiples of each other.
Hence, $\rank(A)\le 1$, $\rank(D)\le 1$, and $\rank(A+D) \le 2$.
Consequently, $\rank(\mathcal{E}_{\text{loss},1}[\rho]) \le 4$.

If we combine several error channels with different probabilities, the density matrix has the form
\begin{equation} \label{eq:channel}
\mathcal{E}[\rho] = p_0\rho + \sum_{i=1}^n \Bigl(
p_i\mathcal{E}_{\text{bit},i}[\rho] 
+q_i \mathcal{E}_{\text{phase},i}[\rho] 
+r_i \mathcal{E}_{\text{loss},i}[\rho] \Bigr),
\end{equation}
where $p_i,q_i,r_i$ are non-negative real numbers adding to $1$.
Noting that the rank of a sum is less than or equal to the sum of the ranks, we obtain that
\begin{equation} \label{eq:channel-rank}
\rank(\mathcal{E}[\rho]) \le (6n+1)\rank(\rho).
\end{equation}
In particular, if we are trying to prepare a pure state $\rho_\text{target}$ with $\rank\rho_\text{target} = 1$, we may instead get a state $\mathcal{E}[\rho_\text{target}]$ with $\rank(\mathcal{E}[\rho_\text{target}]) \le 6n+1$.

\section{Estimating Pauli observables using measurement settings} \label{appen:meas-to-obs}
This appendix provides additional details on the measurement settings described in Section~\ref{subsec:settings-comp}.
We first introduce projectors for the Pauli matrices:
\begin{align}\label{eq:X01}
    X_0 &= \frac{1}{2}\begin{pmatrix}
        1 & 1 \\
        1 & 1
    \end{pmatrix},
    &X_1 &= \frac{1}{2} \begin{pmatrix}
        1 & -1 \\
        -1 & 1
    \end{pmatrix}, \\ \label{eq:Y01}
    Y_0 &= \frac{1}{2} \begin{pmatrix}
        1 & -i \\
        i & 1
    \end{pmatrix},
    &Y_1 &=\frac{1}{2}\begin{pmatrix}
        1 & i \\
        -i & 1
    \end{pmatrix}, \\ \label{eq:Z01}
    Z_0 &= \begin{pmatrix}
        1 & 0 \\
        0 & 0
    \end{pmatrix},
    &Z_1 &= \begin{pmatrix}
        0 & 0 \\
        0 & 1
    \end{pmatrix}.
\end{align}
These matrices satisfy
\begin{align}
    X_0 + X_1 = Y_0 + Y_1 = Z_0 + Z_1 = I
\end{align}
and
\begin{align}
    X_0 - X_1 = X, \\
    Y_0 - Y_1 = Y, \\
    Z_0 - Z_1 = Z.
\end{align}
When we measure according to a measurement setting $S=S_1\dots S_n \in \{X, Y, Z\}^n$, we obtain an outcome distribution $p_S$ over the $2^n$ possible bitstrings: $\{p_S(b)\}_{b \in \{0,1\}^n}$. We have \cite{nielsen_quantum_2012}:
\begin{equation}
    p_S(b) = \Tr\bigl[\bigl((S_1)_{b_1} \otimes \cdots \otimes (S_n)_{b_n}\bigr) \rho^*\bigr],
\end{equation}
where $S_k \in \{X, Y, Z\}$, $b_k\in\{0,1\}$, and $(S_k)_{b_k}$ is defined according to the notation \eqref{eq:X01}--\eqref{eq:Z01} for $1 \le k \le n$.

For $a \in \{0,1\}^n$, we define the Pauli observable $P(S, a) \in \mathcal{P}_n$ by ($1 \le k \le n$):
\begin{equation}
    P(S, a)_k = (S_k)^{a_k} = \begin{cases}
        I, & \text{if } \; a_k = 0, \\
        S_k, & \text{if } \; a_k = 1.
    \end{cases}
\end{equation}
In the following, we provide two examples of how to estimate $\Tr[P(S, a) \rho^*]$ using the outcome distribution $p_S$, and then we formulate the general principle. 

Consider $S = XY$. First, we take $a$ to be the bitstring $11$, so that $P(S, a) = X \otimes Y$. Then we have
\begin{align*}
    \Tr[P(S,a) \rho^*] &= \Tr[(X \otimes Y)\rho^*] \\
    &= \Tr[((X_0 - X_1) \otimes (Y_0-Y_1))\rho^*] \\
    &= \Tr[(X_0 \otimes Y_0) \rho^*] - \Tr[(X_0 \otimes Y_1)\rho^*] \\ &- \Tr[(X_1 \otimes Y_0) \rho^*] + \Tr[(X_1 \otimes Y_1) \rho^*] \\
    &= \bigl(p_{XY}(00) + p_{XY}(11)\bigr) \\ &- \bigl(p_{XY}(01) + p_{XY}(10)\bigr). 
\end{align*}
Note that the positive contributions to $\Tr[(X \otimes Y)\rho^*]$ come from outcome bitstrings with even parity, and negative contributions to $\Tr[(X \otimes Y)\rho^*]$ come from outcome bitstrings with odd parity.

If instead $a = 01$, then $P(S, a) = I \otimes Y$, and
\begin{align*}
    \Tr[P(S,a) \rho^*] &= \Tr[(I \otimes Y)\rho^*] \\
    &= \Tr[((X_0 + X_1) \otimes (Y_0-Y_1))\rho^*] \\
    &= \Tr[(X_0 \otimes Y_0)\rho^*] - \Tr[(X_0 \otimes Y_1)\rho^*] \\ &+ \Tr[(X_1 \otimes Y_0)\rho^*] - \Tr[(X_1 \otimes Y_1)\rho^*] \\
    &= \bigl(p_{XY}(00) + p_{XY}(10)\bigr) \\ &- \bigl(p_{XY}(01) + p_{XY}(11)\bigr). 
\end{align*}
In this case, the positive contributions come from the second bit being $0$ (even parity), and the negative contributions come from the second bit being $1$ (odd parity); the first bit has no effect. Intuitively, this result is expected: measuring $I \otimes Y$ corresponds to not measuring the first qubit, and therefore we marginalize over the outcome of measuring $X$ on the first qubit. It is clear that similar results will hold for $a = 10$ (marginalize over the second qubit) and $a = 00$ (marginalize over both qubits).

These examples suggest how we can estimate $\Tr[P(S, a)\rho^*]$ for any $S$ and $a$. We take the outcome distribution $p_S$ and first marginalize over the qubits $j$ for which $a_j = 0$. Then, for each bitstring $b'$ in the marginal distribution, we compute the parity of $b'$. If the parity is even, we add $p_S(b')$; if it is odd, we subtract $p_S(b')$. We can express these ideas formally as follows. Let $f\colon \{ 0, 1 \}^n \rightarrow \{-1, 1\}$ return $1$ on a bitstring $b \in \{0,1\}^n$ if $b$ has even parity and $-1$ if $b$ has odd parity. Then we have that:
\begin{equation} \label{eq:marginalize-bitstrings}
    \Tr[P(S,a) \rho^*] = \sum_{b \in \{ 0, 1 \}^n} f(b \land a) p_S(b),
\end{equation}
where $\land$ denotes the bitwise AND operation. This bitwise AND operation is equivalent to marginalizing, since if $a_i = 0$ for some $i$, then $b_i \land a_i = 0$. Hence $b_i$ is ignored when $a_i = 0$. On the other hand, when $a_i = 1$, we have $b_i \land a_i = b_i$, and such bits contribute to the parity.

Since~\eqref{eq:marginalize-bitstrings} holds for any $a \in \{0, 1\}^n$, each measurement setting $S$ allows us to compute the expectation values for $2^n$ Pauli observables. However, note that for two measurement settings $S_1, S_2 \in \{X,Y,Z\}^n$, the sets $P_{S_1} = \{P(S_1,a)\colon \, a \in \{0,1\}^n\}$ and $P_{S_2} = \{P(S_2,a)\colon \, a \in \{0,1\}^n\}$ are not disjoint. For example, if $S_1 = XY$ and $S_2 = XZ$, then $P_{S_1} \cap P_{S_2}$ = $\{I \otimes I, X \otimes I\}$.

Similarly to our discussion in Section~\ref{sec:qst}, we do not have access to the true outcome distribution $p_S$; we only have access to the estimate $\widehat{p}_S$ given by the shot data. Hence,
\begin{equation}
    \Tr[P(S,a) \rho^*] \approx \sum_{b \in \{ 0, 1 \}^n} f(b \land a) \widehat{p}_S(b),
\end{equation}
where the approximate equality is due to shot noise.
\end{document}